# A Validated Data-driven Subject and Vehicle Specific Nonlinear Biodynamic Model for Predicting Upper-Body Response in Vehicle Ride

Rahid Zaman, Ahmed Zubayer Raiyan, Md Navid Imtiaz Rifat, Mohammad Ibrahim Hossain, Ashfaq Adnan*

## 1 ABSTRACT:

A six-degree-of-freedom (6-DOF) nonlinear lumped-parameter biodynamic model of the seated human upper body is formulated to predict human response under different unknown loading conditions. The model consists of six anatomically partitioned cascade body segments: pelvis, lower torso, central torso, upper torso, neck, and head. The joints are connected through nonlinear viscoelastic joints incorporating strain-stiffening restoring forces consistent with the nonlinear constitutive behavior of biological soft tissue. Fifteen subject-specific mechanical parameters are considered as design variables. The design variables are six joint natural frequencies, six viscous damping ratios, one Rayleigh coefficient, one nonlinear stiffening scale, and one mass scale. The subject-specific design variables are identified from three distinct vehicular loading cases using a hybrid two-phase optimization strategy. Particle Swarm Optimization (PSO) globally searches on a frequency-domain surrogate and seeds a bounded Nelder-Mead refinement for the full nonlinear Ordinary Differential Equation (ODE) response. A rigorous parameter idealization methodology, based on error-weighted geometric mean in log space, was implemented to fuse the multi-case optima into a single subject-representative consensus parameter set. The idealized model retains physical interpretability across loading scenarios and is subsequently used in forward dynamics to predict and validate the subject's experimental response. Injury risk is assessed using the Dynamic

* Corresponding author: aadnan@uta.edu, ashfaqadnan@gmail.com

Response Index (DRI), the Head Injury Criterion (HIC15), and frequency-weighted RMS acceleration and Vibration Dose Value (VDV). The framework constitutes a complete subject-specific biodynamic pipeline, from experimental data through optimization, idealization, and forward prediction, suitable for occupational health assessment and protective equipment design across diverse loading scenarios.

## 2 INTRODUCTION

The seated upper-body biodynamic response to transient excitation is a critical occupational health concern affecting millions of workers across diverse environments, including vehicular transportation, marine operations, military applications, and different industrial settings. This is especially true for military high-speed boat operators. An increasing number of powerful high-speed boats are in use by sea rescue institutions, coast guards, border patrols, customs teams, law enforcement, and a variety of offshore industries. (Allen et al., 2008; Antao & Soares, 2008; Ullman et al., 2022). Prolonged or high-magnitude exposure has been epidemiologically linked to lumbar spine disorders, intervertebral disc degeneration, chronic low back pain, and neurological impairments such as motion sickness, headache, etc., for air, sea, and land vehicles. Understanding the dynamic mechanical response of the seated upper body is essential for developing effective seat suspension systems, establishing exposure limit values, predicting injury risk, and optimizing occupant protection across these applications. In a study based on the responses of the fast boat operators, it was found that a total of 32% of respondents suffered back-related injuries, 21% of respondents suffered neck-related injuries, 33% reported loss of consciousness, and 70% reported having experienced impaired capacity to perform their job onboard because of impact exposure (Ullman et al., 2022).

Vertical vibration is usually the dominant source causing discomfort, injury, and health disorders. The biodynamic response of the human body under vertical vibration has thus been extensively studied. The seated human body exhibits complex frequency-dependent mechanical behavior characterized by resonance phenomena in the 1–12 Hz range (Kim et al., 2011). Experimental measurements of biodynamic response functions, including apparent mass (AM), seat-to-head transmissibility (STHT), and mechanical impedance, have been extensively documented in controlled laboratory settings using rigid seats and sinusoidal or random vibration inputs. However, direct measurement in real-world transient shock scenarios remains challenging due to experimental and computational complexity, time-consuming procedures, motion artifact contamination, and ethical constraints on human subject exposure to potentially injurious vibration levels. With increasing vibration magnitude, nonlinearity has been reported in apparent mass and transmissibility from vibration at the seat to the head and to various locations along the spine and pelvis (Matsumoto & Griffin, 2002; Nawayseh & Griffin, 2003). Nonlinear phenomena are associated with the properties of soft tissue and the musculoskeletal structures of the human body (Huang & Griffin, 2008; Lakie & Robson, 1988). For long-term low-frequency vibration, the human body shows thixotropic behavior, in which the structure's stiffness decreases with movement (Huang & Griffin, 2008; Lakie & Robson, 1988). However, for sudden changes in vibration or an instantaneous impact, the human body exhibits a strain-stiffening response due to the neuromuscular reflex (Zaman et al., 2025).

Mathematical modeling provides a complementary approach to experimental biomechanics, enabling parametric studies, injury mechanism investigation, and prospective design optimization for seat design, vehicle suspension tuning, marine craft seating, blast mitigation systems, and occupational health risk assessment. Two major factors are important for successful mathematical modeling. One is the dynamic model with representative parameters, and the other is the optimization technique that finds the idealized subject-specific parameters.

Three kinds of dynamic models are widely used to assess or estimate the biodynamic response of seated human body: lumped parameter models (Matsumoto & Griffin, 2001; Nawayseh & Griffin, 2003; Nawayseh & Griffin, 2009; Qiu & Griffin, 2011; Stein et al., 2009), multibody dynamic models (Desai et al., 2018; Desai et al., 2021; Kim et al., 2011; Liang & Chiang, 2008; Wu & Qiu, 2020; Wu et al., 2022; Zanoni et al., 2020; Zheng et al., 2011) and finite element models (Dong et al., 2020; Kitazaki & Griffin, 1997; Kong & Goel, 2003; Liu et al., 2015; Pankoke et al., 1998). Finite element (FE) models are time-consuming and computationally expensive, and it is difficult to match and fit them to experimental acceleration data to find the best parameters for subject-specific cases. Also, the FE model is a redundant system, making it very difficult for multi-segment parameter identification and sensitivity analysis. Multibody models are faster than FE models, especially for spatial kinematics, complex joints, and rotational movements. However, for vertical vibration and acceleration analysis, lumped models with a dominant vertical axis, nonlinear biological properties, and parameter optimization are computationally efficient approaches for generating subject-specific parameter identification.

Lumped-parameter models represent the upper body as a discrete chain of rigid masses interconnected by spring-damper elements, offering computational efficiency and parametric transparency compared to finite element approaches. Multi-degree-of-freedom models incorporating anatomically distinct segments (pelvis, torso subdivisions, neck, head) have demonstrated improved fidelity in predicting accelerations, transmissibility, and apparent mass across the physiological frequency range.

Previous biodynamic models have either neglected nonlinearity entirely or employed population-based polynomial fits with anatomical grounding. Most validation studies employ steady-state sine-wave or broadband random excitation in laboratory settings, which do not represent the transient shocks that dominate injury risk in real operating environments. In addition to focusing on apparent mass and seat-to-head transmissibility, accurately predicting body-segment accelerations is very important, as acceleration is directly correlated with head injury criteria, neck injury criteria, vibration dose value, and motion sickness-related disorders. Very few studies have been conducted to accurately predict segmental human body acceleration. A fundamental limitation of population-averaged biodynamics models is their inability to capture inter-subject variability in resonance frequencies, damping characteristics, and nonlinear stiffening behavior. Population-averaged parameters provide useful information but can misinterpret individual responses in resonance frequency and peak acceleration amplitude.

Subject-specific identification has been explored recently using finite element models, neural network approach (Wang et al., 2026), and different optimization (Alabi et al., 2023; Bae & Kang, 2018; Boileau & Rakheja, 1998; Chen et al., 2011; Desai et al., 2021; Guruguntla et al., 2026; Kim et al., 2005; Valente et al., 2017; Valente et al., 2014; Wang et al., 2026; Zanoni et al., 2020) , but practical deployment has been limited by the high computational cost of time domain optimization and the challenge of selecting globally optimized parameters in a high-dimensional, non-convex parameter space.

Parameter identification in biodynamics model has been approached through manual frequency domain fitting, gradient-based nonlinear least squares (Boileau & Rakheja, 1998; Liang & Chiang, 2006), genetic algorithms (Desai et al., 2018; Desai et al., 2021; Kim et al., 2005; Wu et al., 2022; Zanoni et al., 2020), Firefly algorithm (Guruguntla & Lal, 2021; Guruguntla et al., 2026) and particle swarm optimization (PSO) (Kuang et al., 2023). Gradient-based methods achieve precise local convergence but are sensitive to initialization points and may converge to incorrect local minima. Furthermore, they have expensive finite-difference Jacobian computation for large dimensions. Genetic algorithms provide global search, but are very expensive and time-consuming for Ordinary Differential Equation (ODE)-based objectives (Kim et al., 2011). PSO has excellent global search capability through a combination of cognitive and social exploration mechanisms. However, it is not designed to converge to a precise local minimum. The Nelder-Mead simplex method is faster than gradient methods, with excellent precision near smooth minima, but has no global search capabilities (Nawayseh & Griffin, 2009; Stein et al., 2009). Recently, machine learning and neural networks have been applied to the identification of biodynamic parameters. It has near-instantaneous parameter prediction capability once the surrogate network is trained; however, it requires thousands of ODE evaluations to train the surrogate. Recently, there has been some work on multi-phase optimization for biodynamic models that combines global search reliability with local convergence precision (Wang et al., 2026).

Despite these advances, several critical challenges persist: (1) anthropometric scaling to representative subject characteristics, (2) subject-specific parameter identification, (3) nonlinearity representation capable of capturing strain-stiffening behavior; and (4) experimental validation under realistic transient field conditions. A fundamental challenge in subject-specific biodynamic modeling is parameter identifiability, which requires appropriate representations of a specific human subject's body. However, optimization based on a single loading condition or a similar type of loading condition typically yields parameters that overfit that specific excitation while failing to generalize.

This motivates multi-case identification, in which parameters are simultaneously optimized across multiple distinct loading scenarios, thereby improving the robustness of the identified model via two-phase optimization. This study addresses these challenges by developing and experimentally validating a 6-DOF nonlinear upper-body biodynamic model that features anthropometric scaling, cascade-mass frequency-targeted stiffness identification, hybrid physical-Rayleigh damping, and strain-stiffening nonlinearity calibrated to biological tissue behavior. We adopted a two-phase optimization approach that combines a fast and simplified model with a highly accurate one. Phase 1 deploys PSO on a linearized Frequency Response Function (FRF) surrogate to locate the global minimum basin, and phase 2 applies a bounded Nelder-Mead algorithm to the full ODE nonlinear objective function. This strategy ensures precise local convergence to the true nonlinear minimum. We employed automotive road loading scenarios to test our model in different excitation regimes: (i) continuous road roughness (steady state broadband), (ii) transient pothole impacts with impulsive loading, and (iii) high amplitude railway crossing shocks. No published biodynamic model has been simultaneously calibrated and validated across multiple regimes for individual subjects. This strategy reduces total computation time while retaining global search reliability and full nonlinear fidelity in the final solution.

This paper makes the following contributions: (1) a fully subject-specific 6-DOF nonlinear biodynamic model with anthropometric scaling; (2) a two-phase optimization framework (PSO and bounded Nelder-Mead ODE refinement) validated across two subjects and three loading conditions per subject, (3) a rigorous parameter idealization methodology based on error-weighted geometric mean statistics, producing a consensus subject model from multiple optimization cases, and (4) finally forward dynamics predictions under diverse loading conditions with comprehensive injury risk assessment. We found that this approach reduces model complexity and the computing time required to obtain subject-specific parameters without sacrificing global search reliability or nonlinear precision. Consequently, it enables rapid and personalized injury assessment.

# 3 BIOMECHANICAL MODEL FORMULATION

## 3.1 Model Architecture and Coordinate System

The seated upper body is modeled as a vertically stacked chain of N = 6 rigid body segments connected by nonlinear viscoelastic joints, as illustrated in Figure 1. The segments, ordered from inferior to superior, are: (1) pelvis, (2) lower torso, (3) central torso, (4) upper torso, (5) neck, and (6) head. Each segment possesses a single translational degree of freedom in the vertical direction. The seat pan is treated as a kinematic boundary condition: its measured vertical acceleration $a_{seat}(t)$ provides the forcing input through a base-excitation mechanism, consistent with the standard ISO 5982 measurement protocol in which the accelerometer is located on the hard seat surface beneath any cushioning.

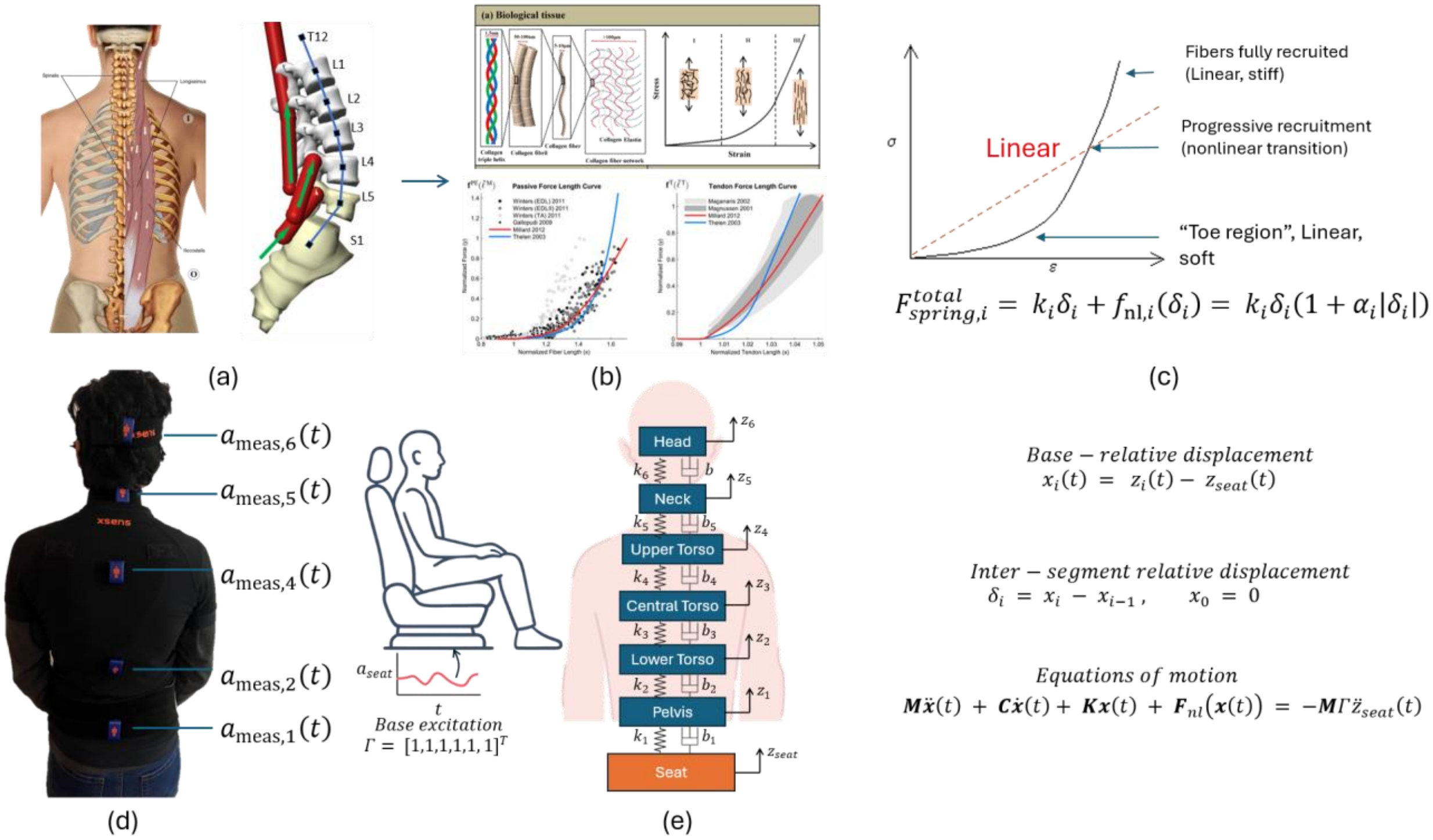

*Figure 1: Schematic of the nonlinear 6-DOF seated upper-body biodynamic model (a) Simplified anatomical model of the spine and nearby muscles;(b) nonlinear behavior of biological tissue, as reported in the literature (Ma et al., 2017; Millard et al., 2013);(c) stress–strain ($\sigma - \varepsilon$) phenomenological curve demonstrating the soft linear toe region, the progressive recruitment nonlinear transition zone, and the fully recruited stiff linear zone governed by the total joint spring force;(d) experimental subject instrumentation layout highlighting the placement of triaxial accelerometers; and (e) assembled lumped-parameter mechanical representation showing the cascaded 6-DOF chain topology with localized joint properties and coordinate systems.*

The ODE is solved in relative (seat frame) coordinates. Let $x_i(t)$ denote the relative displacement of segment $i$ with respect to the seat.

$$x_i(t) = z_i(t) - z_{seat}(t), \quad i = 1, 2, \ldots, 6 \tag{1}$$

where $z_i(t)$ is the absolute vertical displacement of segment $i$ and $z_0(t) = z_{seat}(t)$ is the absolute vertical displacement of the seat. This relative formulation eliminates gravitational drift and simplifies the equations of motion. The absolute acceleration $\ddot{z}_i(t)$ of each segment can be expressed as:

$$\ddot{z}_i(t) = \ddot{x}_i(t) + \ddot{z}_{seat}(t) \tag{2}$$

where $\ddot{z}_{seat}(t)$ is the measured seat acceleration time history (input signal) and $\ddot{x}_i(t)$is the second time derivative of the relative displacement $x_i(t)$ of segment $i$.

### 3.2 Equations of Motion in Relative Coordinates

Applying Newton's second law to each segment in the relative-coordinate frame and assembling into global matrix form, the nonlinear equations of motion under base excitation are:

$$\boldsymbol{M}\ddot{\boldsymbol{x}}(t) + \boldsymbol{C}\dot{\boldsymbol{x}}(t) + \boldsymbol{K}\boldsymbol{x}(t) + \boldsymbol{F}_{nl}(\boldsymbol{x}(t)) = -\boldsymbol{M}\boldsymbol{\Gamma}\ddot{z}_{seat}(t) \tag{3}$$

where $\boldsymbol{M}, \boldsymbol{C},$ and $\boldsymbol{K}$ are the 6×6 global mass, damping, and stiffness matrices, respectively; $\boldsymbol{x}(t), \dot{\boldsymbol{x}}(t)$ and $\ddot{\boldsymbol{x}}(t)$ are the 6×1 base-relative displacement, velocity, and acceleration vectors where $\boldsymbol{x}(t) = [\boldsymbol{x}_1(t), \boldsymbol{x}_2(t), \boldsymbol{x}_3(t), \boldsymbol{x}_4(t), \boldsymbol{x}_5(t), \boldsymbol{x}_6(t)]^{\boldsymbol{T}}$, $\boldsymbol{\Gamma} = [1, 1, 1, 1, 1, 1]^{\mathrm{T}}$ is the unit influence vector coupling base excitation to all DOFs, and $\boldsymbol{F}_{nl}(\boldsymbol{x})$ is the nonlinear strain-stiffening force vector; and $\ddot{z}_{seat}(t)$ is the scalar seat base acceleration.

### 3.3 State-Space Formulation and Numerical Integration

The second-order system Eq (4) is rewritten as a first-order ODE for numerical integration:

$$\dot{\boldsymbol{y}} = \begin{bmatrix} \dot{x}(t) \\ \ddot{x}(t) \end{bmatrix} = \begin{bmatrix} \dot{\boldsymbol{x}}\,(t) \\ \boldsymbol{M}^{-1}(-\boldsymbol{M}\boldsymbol{\Gamma}\ddot{z}_{seat}(t) - \boldsymbol{C}\dot{\boldsymbol{x}}(t) - \boldsymbol{K}\boldsymbol{x}(t) - \boldsymbol{F}_{nl}(\boldsymbol{x}(t))) \end{bmatrix} \tag{4}$$

The state equations are integrated using MATLAB ode45 (explicit Runge–Kutta, relative tolerance $10^{-6}$, absolute tolerance $10^{-9}$, maximum step size $10^{-3}$ s). The raw seat acceleration from the sensor is linearly interpolated to match the numerical solver's time steps.

### 3.4 Segment anatomy and anthropometry

Segmental masses are adopted from the anthropometric database (Boileau & Rakheja, 1998; De Leva, 1996; Winter, 2009), which provides mass fraction coefficients for the major body segments as a function of total body mass and stature. Segment masses ($m_i$) can be found from subject body mass ($M_{body}$):

$$m_i = m_{scale}\,\boldsymbol{\gamma}_i\,M_{body}, \qquad i = 1\,(pelvis), 2, \dots\, 6\,(head) \tag{5}$$

$$\boldsymbol{\gamma} = [0.099, 0.117, 0.162, 0.126, 0.018, 0.059]^T$$

Where, $\gamma_i$ is a mass fraction vector coefficient corresponding to the $i^{th}$ body segment. A single scalar $m_{scale}$ is identified per subject to account for individual upper-body mass distribution variability, reducing 6 mass unknowns to 1. $m_{scale}$ is a design variable in the optimization step. Each DOF supports itself and all segments above. So, the cascade masses ($m_{ca,i}$):

$$m_{ca,i} = \sum_{j=i}^{6} m_j\,, \qquad i = 1,2,\dots,6 \tag{6}$$

The diagonal global mass matrix:

$$\boldsymbol{M} = \operatorname{diag}\left(m_{ca,1}, m_{\mathrm{ca},2}, \dots, m_{\mathrm{ca},6}\right) \in \mathbf{R}^{6\mathrm{x}6} \tag{7}$$

### 3.5 Stiffness Calibration by Cascade Natural-Frequency Assignment

Each joint stiffness $k_i$ is derived from a target natural frequency $f_{n,i}$ and the cascade mass $m_{ca,i}$:

$$k_i = \omega_{\mathrm{n,i}}^2\,\mathrm{m_{ca,i}} = \left(2\pi\, f_{\mathrm{n},i}\right)^2\, m_{\mathrm{ca},i} \quad i = 1, 2, \dots, 6 \tag{8}$$

The natural frequencies ($f_{\mathrm{n},i}$) are design variables in the optimization phase. The global stiffness matrix, $\boldsymbol{K}$**,** has the standard tridiagonal topology for a chain system, where the diagonal entries are:

$$\boldsymbol{K}_{ii} = k_i + k_{i+1}, \qquad \text{for i} = 1, \dots, 5 \tag{9}$$

$$\boldsymbol{K}_{66} = k_6$$

And the symmetric off-diagonal entries are defined by:

$$\boldsymbol{K}_{i,i+1} = \boldsymbol{K}_{i+1,i} = -k_{i+1}, \qquad for\ i = 1, \dots, 5$$

with the boundary condition $k_0 = 0$ (seat end fixed), and $k_7 = 0$ (head, free end).

### 3.6 Damping Matrix Construction

The global damping matrix is constructed as a two-component superposition:

$$\boldsymbol{C} = \boldsymbol{C}_{\text{phys}} + \boldsymbol{C}_{\text{Ray}} \alpha_r \boldsymbol{M} \tag{10}$$

The physical component, $\mathbf{C}_{phys}$, has the same tridiagonal topology as $\boldsymbol{K}$, with joint damping coefficients $b_i$ replacing with the joint stiffness coefficients $k_i$ to represent inter-segment viscous dissipation. These coefficients are determined from the critical damping ratios, $\zeta_i$, assigned to each joint using the cascade-mass formula:

$$b_i = 2\,\zeta_i \sqrt{k_i m_{\text{ca},i}} \tag{11}$$

The damping ratios ($\zeta_i$) are considered design variables in this study.

The Rayleigh mass-proportional component adds supplementary modal damping via the scalar coefficient:

$$\boldsymbol{C}_{\text{Ray}} = \alpha_r \boldsymbol{M} \tag{12}$$

The modal damping contribution from Rayleigh damping in mode $p$ with natural frequency $\omega_{n,r}$ $is$:

$$\zeta_p = \frac{\alpha_r}{2\,\omega_{n,p}} \tag{13}$$

In this study, $\alpha_r$ is a design variable that is optimized in PSO and Nelder-Mead with bounds. After optimization, the total damping ratio for mode $p$ is verified using the modal projection:

$$\zeta_p = \frac{\boldsymbol{\varphi}_p{}^{\mathrm{T}} \boldsymbol{C} \boldsymbol{\varphi}_p}{2\,\omega_{n,p}\, \boldsymbol{\varphi}_p{}^{\mathrm{T}} \boldsymbol{M}\, \boldsymbol{\varphi}_p} \tag{14}$$

where $\boldsymbol{\varphi}_p$ is the mass-normalized eigenvector satisfying $\varphi_p{}^{\mathrm{T}}\boldsymbol{M}\,\varphi_p = 1$ (Inman & Singh, 1994; Rao, 2017).

### 3.7 Nonlinear Constitutive Relations for the Intervertebral Joints

Each viscoelastic joint represents the combined mechanical response of the intervertebral disc, surrounding ligamentous complex, and adjacent soft tissue. Under axial compression, spinal soft tissue exhibits a J-shaped force–displacement curve, as shown in Figure 1(c), with an initial low-stiffness "toe region" followed by progressive stiffening as collagen fibers are recruited (Ma et al., 2017; Panjabi et al., 1994; Shirazi-Adl et al., 1984). This behavior is captured by a nonlinear restoring force $f_{\mathrm{nl},i}(\delta_i)$ acting inside segment $i$:

$$f_{\mathrm{nl},i}(\delta_i) = k_i \alpha_{\mathrm{i}} \delta_i |\delta_i| \tag{15}$$

where $k_i$ is the linear stiffness at segment $i$, $\delta_i$ is inter-segment displacement at segment $i$ and $\alpha_i$ $[m^{-1}]$ is the strain-stiffening coefficient.

$$\alpha_i = s_{nl}\bar{\alpha}_i \tag{16}$$

Where, $s_{nl}$ is a single global nonlinear-stiffening scale factor, identified once per subject as a design variable, and $\bar{\alpha}_i$ is a nominal per-joint strain-stiffening shape vector. To isolate the structural deformation of the individual joints from the global kinematics, we define the inter-segment relative displacement $\delta_i$ as:

$$\delta_i = x_i - x_{i-1}, \qquad x_0 = 0$$

The total spring force at segment $i$ including linear and nonlinear part is:

$$F_{spring,i}^{total} = k_i\delta_i + f_{\mathrm{nl},i}(\delta_i) = k_i\delta_i(1 + \alpha_i|\delta_i|) \tag{17}$$

The linear part $k_i\delta_i$ is captured by $\boldsymbol{Kx}(t)$ and nonlinear part $k_i\alpha_i\delta_i|\delta_i|$ is captured by $\boldsymbol{F_{nl}}(\boldsymbol{x}(\boldsymbol{t}))$ in Eq (3). The nonlinear joint force $f_{nl,i}$ acts downward on DOF $i$ and upward on DOF $(i-1)$. Assembly into the global DOF force vector:

$$\boldsymbol{F}_{nl}(\boldsymbol{x}(\boldsymbol{t})) = \begin{bmatrix} (F_{nl})_1 \\ (F_{nl})_2 \\ \vdots \\ (F_{nl})_6 \end{bmatrix}, \quad where(\boldsymbol{F_{nl}})_i = \begin{cases} f_{nl,i} - f_{nl,i+1}, & i = 1,2,.,5 \\ f_{nl,6}, & i = 6 \end{cases} \tag{18}$$

This assembly guarantees that every joint force is shared equally and oppositely between two segments it connects, and consistent with Newton's third law.

# 4 BIODYNAMIC RESPONSE AND INJURY CRITERIA

## 4.1 Apparent mass and Mechanical Impedance

The driving-point apparent mass $\mathrm{AM}(f)$ is estimated as the seat reaction force $F_{\mathrm{seat}}(f)$ divided by seat acceleration$a_{\mathrm{seat}}(f)$:

$$\mathrm{AM}(f) = \frac{F_{\mathrm{seat}}(f)}{a_{\mathrm{seat}}(f)} \tag{19}$$

The seat reaction force in the time domain is computed from the relative displacement $x_1(t)$ and relative velocity of the pelvis $\dot{x}(t)$:

$$F_{\mathrm{seat}}(t) = K_{11}\, x_1(t) + C_{11}\, \dot{x}_1(t)$$

Where $K_{11}$ *and* $C_{11}$ are the (1,1) entries of the global stiffness and damping matrices, representing the total stiffness and damping exerted on the pelvis from the seat. $\mathrm{AM}(f)$ is estimated via cross-spectral density, H1 estimator using Welch's method:

$$\mathrm{AM}(f) = \frac{S_{FA}(f)}{S_{AA}(f)}$$

Where $S_{FA}(f)$ is the cross-power spectral density between seat force and seat acceleration and $S_{AA}(f)$ is the auto power spectral density of seat acceleration. Finally, it is normalized by the total body mass ($M_{body}$)

$$\overline{\mathrm{AM}}(f) = \frac{|\mathrm{AM}(f)|}{M_{body}} \tag{20}$$

ISO 5982 corridor specifies $0.80 \leq \overline{\mathrm{AM}}^{\mathrm{peak}} \leq 0.95$ for automotive seated subjects on smooth roads. Apparent mass phase angle $\phi_{ap}\,(f)$:

$$\phi_{ap}\,(f) = \arg\left(S_{FA}(f)\,\left(\frac{180°}{\pi}\right)\right)$$

Where $\phi_{ap}\,(f)$ is in degrees and it tracks the physical lag between seat acceleration and the body's force response. A positive phase angle indicates elastic spring forces dominate; a negative phase angle indicates damping and inertial lags dominate; and a sharp drop indicates structural resonance points.

Coherence $(\gamma^2(f))$ between seat force and seat acceleration assesses transfer function statistical reliability:

$$\gamma^2(f) = \frac{|S_{FA}(f)\,|^2}{S_{AA}(f)\,S_{FF}(f)}$$

Where $S_{FA}(f)$ is the cross-power spectral density between seat force and seat acceleration and $S_{AA}(f)$ is the auto power spectral density of seat acceleration, and $S_{FF}(f)$ is the auto power spectral density of seat force. It measures how linearly related two signals are at each frequency and whether the output at that frequency is actually caused by the input or by noise. A value of $\gamma^2(f) > 0.8$ means roughly 80% of the output power is linearly attributed to the input.

### 4.2 Seat-to-head transmissibility and mechanical impedance

The seat-to-head transmissibility ($STHT$) between seat acceleration and head acceleration assesses the vibration transmission through the spine:

$$\mathrm{STHT}(f) = \frac{S_{HA}(f)}{S_{AA}(f)} \tag{21}$$

$S_{HA}(f)$ is the cross-power spectral density between head acceleration and seat acceleration, and $S_{AA}(f)$ is the auto-power spectral density of seat acceleration. It measures how much of the input vibration amplitude is amplified or attenuated by the human body at each frequency.

The driving-point mechanical impedance, $Z_{mech}(f)$ between seat reaction force and seat velocity relates the force response to the input velocity:

$$Z_{mech}(f) = \frac{S_{FV}(f)}{S_{VV}(f)} \tag{22}$$

Where $S_{FV}(f)$ is the cross-power spectral density between seat reaction force and seat velocity, and $S_{VV}(f)$ is the auto-power spectral density of seat velocity. Seat velocity in the time domain is obtained by numerically integrating the seat acceleration using the cumulative trapezoidal rule. Driving-point mechanical impedance measures the dynamic resistance of the seated body structure to the input velocity excitation at each frequency.

### 4.3 Head Injury Criterion ($HIC_{15}$)

Head Injury Criteria (HIC) is proposed by the US National Highway Traffic Safety Administration (NHTSA), as shown in Eq. (23)

$$HIC = max\left[\frac{1}{t_2-t_1}\int_{t_1}^{t_2} a_{head}^{abs}(t)dt\right]^{2.5}(t_2-t_1) \tag{23}$$

Where,$a_{head}^{abs}(t)$ is the absolute head acceleration including gravity, $t_1$ and $t_2$ represent two arbitrary time points during the acceleration pulse. Acceleration is measured in gravitational force ($g$), and time is measured in seconds. $HIC_{36}$ means that $HIC$ is measured for each 36ms apart ($t_2$ and $t_1$ is less than 36 ms apart). Suggested human tolerance for 50th percentile males are: $HIC_{15}$ < 700 (moderate injury boundary); $HIC_{15}$ < 1000 (severe boundary).

### 4.4 Dynamic Response Index (DRI)

The DRI assesses spinal compressive injury under vertical shock (Payne, 1965) modeled as:

$$\ddot{z}_{DRI} + 2\zeta_{DRI}\omega_{DRI}\dot{z}_{DRI} + {\omega_{DRI}}^2 z_{DRI}(t) = -\ddot{z}_{seat}(t), \tag{24}$$

$$\mathrm{DRI} = \frac{{\omega_{DRI}}^2 \, max_t \, |z_{DRI}(t)|}{g} \tag{25}$$

Where $\omega_{DRI} = 52.9\frac{rad}{s}$, $\zeta_{DRI} = 0.224$ for spinal damping ratio, $z_{DRI}$ is the spinal compression response. Risk categories: DRI $<$ 5.2 (low), 5.2–13.0 (medium), 13.0–17.7 (high), and > 17.7 (critical/not survivable). DRI =17.7 consider of AIS 2+ spinal injury (Thyagarajan et al., 2014).

### 4.5 Frequency-Weighted RMS Acceleration

The frequency-weighted acceleration ($a_w$) assesses chronic WBV health risk. The acceleration data passed through weighted band pass filter to get the weighted accelerations.

$$a_{w_f} = \left[\int W_k^2(f) S_{AA}(f) df\right]^{1/2} \tag{26}$$

Where $W_k(f)$ is the filter is defined by its frequency dependent weight based on ISO 2631-1. $S_{AA}(f)$ is the power spectral density of acceleration signal. According to EU Directive, The daily action (warning) value for frequency weighted RMS acceleration is 0.5 $m/s^2$ and daily limit value is 1.15 $m/s^2$ (Pelmear & Leong, 2002). $a_{w_f}$ reflects the fact that human body is not equally sensitive to vibration at all frequencies. $W_k$ filter emphasizes the human body's natural frequency where vibration is most harmful to lumbar spine.

### 4.6 Vibration Dose value (VDV)

The VDV is more sensitive to transient peaks (shocks)

$$VDV = \left(\int_0^T a_w^4(t)dt\right)^{\frac{1}{4}} \tag{27}$$

Where $a_w(t)$ is the instantaneous frequency weighted acceleration time history. Daily action limits for VDV is $8.5 m/s^{1.75}$ and daily limit value is 17 $m/s^{1.75}$ (Eger et al., 2013). VDV captures cumulative dose including impulsive contributions. The fourth-power integral makes it sensitive to peak events by making a short-duration spike contribute more than a sustained low-level vibration. This makes VDV particularly relevant for land, sea, and air vehicles injury analysis.

## 5 EXPERIMENTAL METHODOLOGY

### 5.1 Subjects and Protocol

Two adult male subjects participated in the data-collection campaign under an institutionally approved protocol. Subject 1 has a body mass of 58.6 kg and a stature of 1.73 m, yielding an upper-body mass of 34.05 kg from the mass-fraction sum (58.1 % of body mass). Subject 2 has a body mass of 74.0 kg and a stature of 1.70 m, yielding an upper-body mass of 43.0 kg. Both subjects were instructed to maintain an upright seated posture with feet supported, hands resting on the thighs, and no contact with the backrest.

### 5.2 Instrumentation

Six Blue Trident dual-range tri-axial inertial measurement units (IMU) were placed at the seat pan and at five body landmarks: pelvis (S2 level), lower torso (L3 level), upper torso (T6 level), neck (C6 level), and head (over the occipital bone, secured with an elastic headband). Each IMU samples the triaxial acceleration on a low-g channel (16g, 12-bit). The native sampling rate is 1129.3 Hz. Raw IMU signals were pre-processed with a 4th-order Butterworth low-pass filter at 100 Hz cutoff frequency to suppress aliasing artifacts and high-frequency accelerometer noise above the physiologically relevant vibration band. Signal offsets due to gravitational projection ($g \cdot \cos\theta$, where $\theta$ is the sensor tilt angle from vertical) were removed by demeaning all channels before any analysis.

### 5.3 Loading cases:

Three distinct vehicular loading scenarios were used for parameter identification and forward-simulation validation. For all cases, multiple repetitions were also retained for the cross-repetition consistency.

Case 1: Regular road condition

The subject was seated as a passenger during a regular road condition at vehicle speed around 25mph. The sear acceleration exhibits a broadband spectrum with 0.2 to 1 g magnitude. Each data acquisition was recorded three times. The analysis window is around 5s. This case provides data for characterizing frequency-dependent stiffness and damping in the linear operating regime.

Case 2 Road pothole traversal

The subject was seated as a passenger during a deliberate pothole traversal at moderate speed. It generates a transient acceleration pulse of about 1g to 3g acceleration peaks. The analysis window is approximately 4s to 5s in duration. Each loading case was repeated 3 times per subject. The repetition with the highest head-segment RMS acceleration was selected as the representative trial for optimization.

Case 3 Railway grade crossing:

The subject was seated as a passenger in a ground vehicle traversing an at-grade railway crossing (paved road interrupted by two parallel steel rails embedded in asphalt). The crossing generates a sequence of high amplitude impacts with about 3g to 8g accelerations peaks. The analysis window is 4–6 s in duration. Signal Processing

All acceleration channels were filtered with a fourth-order zero-phase Butterworth low-pass filter at a 100 Hz cutoff to suppress high-frequency sensor noise and aliasing. The first 2 s of each ODE integration window constitute a warm-up interval that is excluded from the optimization error metric.

# 6 PARAMETER IDENTIFICATION FRAMEWORK

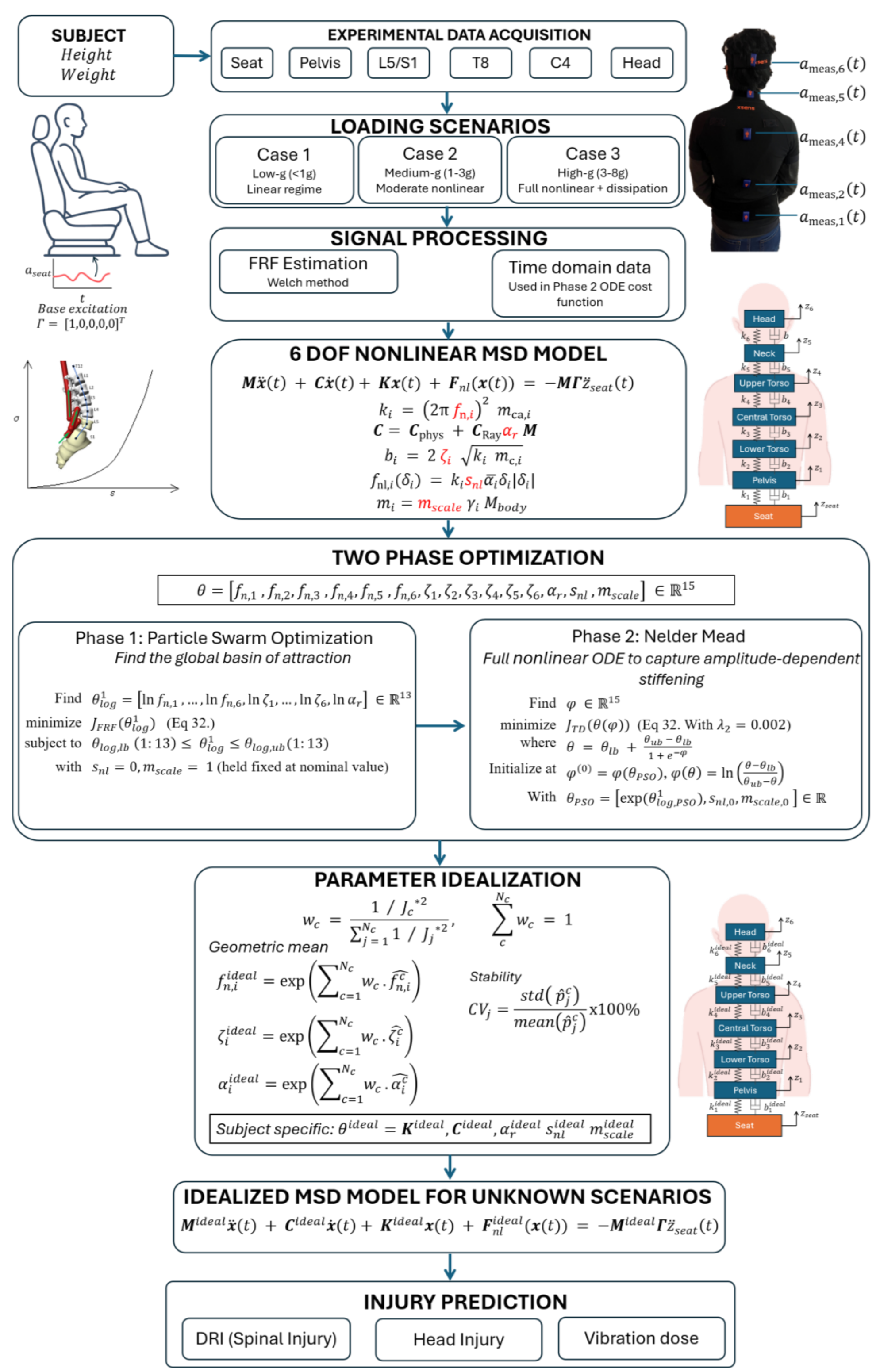

*Figure 2: Comprehensive layout of the experimental data acquisition, dual-phase optimization framework, and injury prediction pipeline for the nonlinear 6-DOF upper-body model.*

### 6.1 Design variables and bounds

Fifteen design variables are simultaneously identified per subject per loading case: six joint natural frequencies $f_{n,i}$ , six joint damping ratios $\zeta_i$ , one global Rayleigh coefficient $\alpha_r$, one nonlinear-stiffening scale $s_{nl}$ , and one upper-body mass-scale factor $m_{scale}$. All optimization variables are represented in log-space during the search to enforce positivity, to normalize the search step across multiple decades, and to ensure log-normal idealization:

$$\theta_{log} = ln\ (\theta) = \begin{bmatrix} \ln(f_{n,1}) \\ \cdot \\ \vdots \\ \ln(f_{n,6}) \\ \ln(\zeta_1) \\ \cdot \\ \cdot \\ \cdot \\ \ln(\zeta_6) \\ \ln(\alpha_r) \\ \ln(s_{nl}) \\ \ln(m_{scale}) \end{bmatrix} \tag{28}$$

Where $\theta$ is the physical parameter vector, $\theta = [f_{n,1}, f_{n,2}, f_{n,3}, f_{n,4}, f_{n,5}, f_{n,6}, \zeta_1, \zeta_2, \zeta_3, \zeta_4, \zeta_5, \zeta_6, \alpha_r\ s_{nl}\ m_{scale}] \in \mathbb{R}^{15}$.

Physical bounds based on the biomechanical literature (Mansfield, 2004; Rao, 2017; Seidel, 2005) are:

*Table 1: Optimization bound from literature*

| Variable | Lower | Upper | Units | Rationale |
|---|---|---|---|---|
| $\boldsymbol{f_{n,1}}$ (seat–pelvis) | 5 | 15 | Hz | Pelvis–seat contact stiffness |
| $\boldsymbol{f_{n,2}}$ (pelvis–lt) | 5 | 18 | Hz | Lumbar L4-S1 elastic resistance |
| $\boldsymbol{f_{n,3}}$ (lt–ct) | 3 | 15 | Hz | Thoracolumbar T6-L3 region |
| $\boldsymbol{f_{n,4}}$ (ct–ut) | 3 | 15 | Hz | Upper thoracic T1-T5 region |
| $\boldsymbol{f_{n,5}}$ (ut–neck) | 5 | 20 | Hz | Cervical C7-T1 transition |

| $\boldsymbol{f_{n,6}}$ (neck–head) | 3 | 12 | Hz | Atlanto-occipital articulation |
|---|---|---|---|---|
| $\boldsymbol{\zeta_i}$ (all six) | 0.025 | 0.40 | — | Spinal viscoelastic damping range |
| $\boldsymbol{\alpha_r}$ | 0.1 | 3 | rad·s$^{-1}$ | Rayleigh modal damping floor |
| $\boldsymbol{s_{nl}}$ | 0.01 | 2.0 | — | Strain-stiffening amplitude scale |
| $\boldsymbol{m_{Scale}}$ | 0.85 | 1.15 | — | Total upper-body mass adjustment |

## 6.2 Objective Function

### 6.2.1 Phase 1: Particle swarm optimization

A particle swarm (80 particles, 200 maximum iterations) searches the 13-dimensional log-space parameters domain using the linearized FRF surrogate objective. The PSO identifies the correct basin of attraction and seeds Phase 2. The optimization structure of the phase 1 surrogate model is:

$$\begin{aligned}
\text{Find} \quad & \theta_{log}^{1} = \left[\ln f_{n,1}, \ldots, \ln f_{n,6}, \ln \zeta_1, \ldots, \ln \zeta_6, \ln \alpha_r\right] \in \mathbb{R}^{13} \\
\text{minimize} \quad & J_{FRF}(\theta_{log}^{1}) \quad \text{(Eq 32.)} \qquad\qquad (29) \\
\text{subject to} \quad & \theta_{log,lb}\,(1{:}13) \leq \theta_{log}^{1} \leq \theta_{log,ub}(1{:}13) \\
\text{with} \quad & s_{nl} = 0, m_{scale} = 1 \quad \text{(held fixed at nominal value)}
\end{aligned}$$

Phase 1 minimizes the weighted sum of frequency-response errors across all instrumented segments and all loading cases. The linearized frequency-domain model ($s_{nl} = 0$) at segment $i$ and frequency $f$ is

$$H_{\text{model},i}(f,\theta) = \boldsymbol{e}_i^T[-(2\pi f)^2\boldsymbol{M} + j(2\pi f)\boldsymbol{C} + \boldsymbol{K}]^{-1}(-\boldsymbol{M\Gamma}) \qquad (30)$$

Where $\boldsymbol{e}_i^T$ is je unit selection vector to extract the $i^{th}$ DOF data from the model, $-\boldsymbol{M\Gamma}$ represent force input vector (base excitation projected on the mass matrix), $j$ is the imaginary unit.

The objective function of PSO phase 1 is:

$$J_{\mathrm{FRF}}(\theta) = \sum_{i}^{6} \sum_{f} \gamma_{\mathrm{x}i}{}^{2}(f)\left[\left|H_{\mathrm{model},i}(f,\theta)\right| - \left|H_{\mathrm{meas},i}(f)\right|\right]^{2} + \lambda_1 \left\|\theta_{\log} - \theta_{\mathrm{ref}}\right\|^{2} \quad (31)$$

Where $\lambda_1 = 0.005$. Segment weights: Pelvis = 0.5, Lower Torso = 0.2, Upper Torso = 0.8, Neck = 2.5, Head = 3.0 (head and neck weighed highest due to injury relevance). Tikhonov regularization strength $\lambda_1 = 0.005$ prevents over-fitting to the Phase 1 FRF surrogate.

### 6.2.2 Phase 2: Bounded time domain ODE refinement

Starting from the PSO result, bounded Nelder-Mead (MATLAB fminsearch) refines all 15 parameters using the full nonlinear ode45 integration (Nawayseh & Griffin, 2009). The optimization structure of Phase 2 is:

$$\begin{aligned} &\text{Find} && \varphi \in \mathbb{R}^{15} \\ &\text{minimize} && J_{TD}(\theta(\varphi)) \quad \text{(Eq 32. With } \lambda_2 = 0.002\text{)} \\ &\text{where} && \theta = \theta_{lb} + \frac{\theta_{ub} - \theta_{lb}}{1 + e^{-\varphi}} \\ &\text{Initialize at} && \varphi^{(0)} = \varphi(\theta_{PSO}),\ \varphi(\theta) = \ln\left(\frac{\theta - \theta_{lb}}{\theta_{ub} - \theta}\right) \\ &\text{With} && \theta_{PSO} = \left[\exp(\theta^{1}_{log,PSO}), s_{nl,0}, m_{scale,0}\right] \in \mathbb{R} \end{aligned} \quad (32)$$

The objective function of time-domain phase 2 is:

$$J_{\mathrm{TD}}(\theta) = \sum_{i}^{6} \frac{\left\|\ddot{x}_{\mathrm{model},i}(t,\theta) - a_{\mathrm{meas},i}(t)\right\|^{2}}{\left\|a_{\mathrm{meas},i}(t)\right\|^{2}} + \lambda_2 \|\theta - \theta_{PSO}\|^{2} \quad (33)$$

Where $\lambda_2 = 0.002$.

Bounds are enforced via smooth sigmoid transformation:

$$\theta = \theta_{lb} + \frac{\theta_{ub} - \theta_{lb}}{1 + e^{-\varphi}},\ \varphi \in (-\infty, +\infty) \quad (34)$$

where $\varphi \in (-\infty, +\infty)$ is the unconstrained search variable. This sigmoid mapping eliminates clamping artifacts present in projection-based methods.

# 7 PARAMETER IDEALIZATION

Parameter idealization aggregates multi-case results into a single consensus estimate that best represents the subject's intrinsic mechanical properties and is suitable for forward predictions under novel loading conditions without re-identification.

## 7.1 Error weighted geometric mean

During the mean calculation, cases with lower residuals get higher weight. The case weights are defined as:

$$w_c = \frac{1 / {J_c}^{*2}}{\sum_{j=1}^{N_c} 1 / {J_j}^{*2}}, \quad \sum_c^{N_c} w_c = 1 \tag{35}$$

Where $J_c$ is the corresponding fitting error achieved by optimization for case $c$, $w_c$ is the weight assigned to optimization case for idealization, $N_c$ is the total number os optimization cases available for this subject, Squaring the inverse error gives stronger preference to better-fitting cases, consistent with weighted least-squares inverse problems (Aster et al., 2018). The idealized parameter is the error-weighted geometric mean in log space which is applied to each parameter independently:

$$p^{ideal} = \exp(\sum_{c=1}^{N_c} w_c \cdot \widehat{p_c}) \tag{36}$$

$$f_{n,i}^{ideal} = \exp\left(\sum_{c=1}^{N_c} w_c \cdot \widehat{f_{n,\iota}^{c}}\right) \tag{37}$$

Where $p^{ideal}$is the idealized  value of parameter $p$, $\widehat{p_c}$ is the optimized value of parameter p from case c,

## 7.2 Parameter reliability Coefficient of Variation

The Coefficient of Variation ($CV \ = \frac{\sigma}{\mu}$ in physical space) quantifies case-to-case parameter scatter:

$$CV_j = \frac{std(\ \hat{p}_j^c)}{mean(\hat{p}_j^c)}\text{x}100\% \tag{38}$$

CV quantifies how consistently a parameter is identified across different loading conditions. A low CV (e.g., for $\alpha_r$, CV ≈ 1–2%) means the parameter is robustly observable regardless of excitation type. It indicates high confidence in its physical value. A high CV (e.g., for $\zeta_i$, CV > 100%) means the parameter is sensitive to the specific loading scenario, likely because different cases excite different frequency bands with different damping observability. CV > 30% is the actionable flag. It indicated poorly constrained and  additional loading cases are needed before that parameter can be reliably idealized.

Finally the idealized models metrices are rebuilt from the consensus parameters

$$k_i^{ideal} = \left(2\pi f_{n,i}^{ideal}\right)^2 . m_{ca,i} \tag{39}$$

$$b_i^{ideal} = 2\zeta_i^{ideal} m_{ca,i} \tag{40}$$

$$\boldsymbol{C}^{ideal} = \boldsymbol{C}_{phy}^{ideal} + \alpha_r^{ideal}\boldsymbol{M} \tag{41}$$

# 8 RESULTS AND VALIDATION

### 8.1 Optimization and idealization results for Subject 1:

Subject 1 has a body mass of 58.6 kg and a stature of 1.73 m. Segmental upper-body masses are: pelvis 5.797 kg, lower 6.852 kg, central 9.487 kg, upper 7.378 kg, neck 1.054 kg, head 3.479 kg (total 34.05 kg).

Table 2 summarizes the optimized parameters and per-segment RMS errors for Subject 1 under the three loading cases.

*Table 2: Optimized parameters for Subject 1*

| Parameter | | Case 1 | Case 2 | Case 3 | Idealized |
|---|---|---|---|---|---|
| $f_{n,i}$ | $f_{n,1}$ (Pelvis) Hz | 14.998 | 14.999 | 13.277 | 14.671 |
| | $f_{n,2}$ (LT) Hz | 17.998 | 10.901 | 5.219 | 11.617 |
| | $f_{n,3}$ (CT) Hz | 3.633 | 3.794 | 14.402 | 4.748 |
| | $f_{n,4}$ (UT) Hz | 11.314 | 10.332 | 14.860 | 11.435 |
| | $f_{n,5}$ (Neck) Hz | 11.461 | 19.998 | 19.812 | 16.044 |
| | $f_{n,6}$ (Head) Hz | 10.348 | 9.815 | 11.890 | 10.375 |
| $\zeta_i$ | $\zeta_1$ (Pelvis) | 0.4000 | 0.4000 | 0.0570 | 0.2812 |
| | $\zeta_2$ (LT) | 0.4000 | 0.4000 | 0.3927 | 0.3986 |
| | $\zeta_3$ (CT | 0.4000 | 0.4000 | 0.0491 | 0.2737 |
| | $\zeta_4$ (UT) | 0.0498 | 0.0486 | 0.0629 | 0.0514 |
| | $\zeta_5$ (Neck) | 0.0500 | 0.1298 | 0.0541 | 0.0761 |
| | $\zeta_6$ (Head) | 0.3223 | 0.0500 | 0.3932 | 0.1509 |
| $\alpha_{nl,i}$ | $\alpha_{nl}$ (Pelvis) ($m^{-1}$) | 8.999 | 4.500 | 7.649 | 6.503 |
| | $\alpha_{nl}$ (LT) ($m^{-1}$) | 7.499 | 3.750 | 6.374 | 5.419 |
| | $\alpha_{nl}$ (CT) ($m^{-1}$) | 5.999 | 3.000 | 5.100 | 4.335 |
| | $\alpha_{nl}$ (UT) ($m^{-1}$) | 4.500 | 2.250 | 3.825 | 3.251 |
| | $\alpha_{nl}$ (Neck) ($m^{-1}$) | 2.400 | 1.200 | 2.040 | 1.734 |
| | $\alpha_{nl}$ (Head) ($m^{-1}$) | 1.200 | 0.600 | 1.020 | 0.867 |
| $nl_{scale}$ | | 0.300 | 0.150 | 0.255 | 0.217 |
| $\alpha_r$ $(rad/s)$ | | 2.9997 | 2.9997 | 2.9330 | 2.9875 |
| $m_{scale}$ | | 1.0001 | 1.0000 | 1.0105 | 1.0019 |
| $best_{err}$ (%) | | 7.91 | 7.59 | 11.65 | 8.45 |

## 8.2 Optimization and idealization results for Subject 2:

Subject 2 has a body mass of 74.0 kg and a stature of 1.70 m. Segmental upper-body masses are: pelvis 7.319 kg, lower 8.649 kg, central 11.977 kg, upper 9.314 kg, neck 1.331 kg, head 4.393 kg (total 43.0 kg).

Table 3 summarizes the optimized parameters and per-segment RMS errors for Subject 1 under the three loading cases.

*Table 3: Optimized parameters for Subject 2*

| Parameter | | Case 1 | Case 2 | Case 3 | Idealized |
|---|---|---|---|---|---|
| $f_{n,i}$ | $f_{n,1}$ (Pelvis) Hz | 10.240 | 14.895 | 14.127 | 12.504 |
| | $f_{n,2}$ (LT) Hz | 16.481 | 5.062 | 17.998 | 13.874 |
| | $f_{n,3}$ (CT) Hz | 3.733 | 14.832 | 5.717 | 5.686 |
| | $f_{n,4}$ (UT) Hz | 6.853 | 14.618 | 12.380 | 10.013 |
| | $f_{n,5}$ (Neck) Hz | 16.533 | 17.881 | 19.998 | 18.140 |
| | $f_{n,6}$ (Head) Hz | 9.597 | 10.149 | 11.999 | 10.633 |
| $\zeta_i$ | $\zeta_1$ (Pelvis) | 0.0342 | 0.0430 | 0.0503 | 0.0418 |
| | $\zeta_2$ (LT) | 0.2273 | 0.2288 | 0.0499 | 0.1213 |
| | $\zeta_3$ (CT | 0.3929 | 0.3362 | 0.4000 | 0.3851 |
| | $\zeta_4$ (UT) | 0.0373 | 0.0356 | 0.0501 | 0.0418 |
| | $\zeta_5$ (Neck) | 0.2573 | 0.3065 | 0.0788 | 0.1624 |
| | $\zeta_6$ (Head) | 0.0563 | 0.3930 | 0.3707 | 0.1735 |
| $\alpha_{nl,i}$ | $\alpha_{nl}$ (Pelvis) ($m^{-1}$) | 29.221 | 1.815 | 18.091 | 14.653 |
| | $\alpha_{nl}$ (LT) ($m^{-1}$) | 24.351 | 1.512 | 15.076 | 12.211 |
| | $\alpha_{nl}$ (CT) ($m^{-1}$) | 19.480 | 1.210 | 12.060 | 9.769 |
| | $\alpha_{nl}$ (UT) ($m^{-1}$) | 14.610 | 0.907 | 9.045 | 7.327 |
| | $\alpha_{nl}$ (Neck) ($m^{-1}$) | 7.792 | 0.484 | 4.824 | 3.907 |
| | $\alpha_{nl}$ (Head) ($m^{-1}$) | 3.896 | 0.242 | 2.412 | 1.954 |
| $nl_{scale}$ | | 0.974 | 0.061 | 0.603 | 0.488 |
| $\alpha_r$ (rad/s) | | 2.298 | 2.936 | 1.315 | 1.904 |
| $m_{scale}$ | | 0.9272 | 1.1237 | 1.0001 | 0.9921 |
| $best_{err}$ (%) | | 7.69 | 11.69 | 7.63 | 8.37 |

### 8.3 Idealized model and Parameter reliability

Table 4 and Table 5 summarize the final subject-specific natural frequencies, damping ratios, linear stiffnesses, nonlinear coefficients, Rayleigh damping, mass scaling, and weighted idealization error that define the validated model.

*Table 4: Idealized parameters for subject 1*

| Joint | $f_n(Hz)$ | $\zeta$ | $k(kN/m)$ | $\alpha_{nl}(m^{-1})$ | Mass $(kg)$ |
|---|---|---|---|---|---|
| Pelvis | 14.671 | 0.2812 | 289.31 | 6.503 | 5.80 |
| Lower torso | 11.617 | 0.3986 | 150.51 | 5.419 | 6.85 |
| Central torso | 4.748 | 0.2737 | 19.04 | 4.335 | 9.49 |
| Upper torso | 11.435 | 0.0514 | 61.48 | 3.251 | 7.38 |
| Neck | 16.044 | 0.0761 | 46.06 | 1.734 | 1.05 |
| Head | 10.375 | 0.1509 | 14.78 | 0.867 | 3.48 |
| Rayleigh $\alpha_r = 2.9875\frac{rad}{s}$, Mass scale =1.002, $nl_{scale} = 0.217$, weighted error= 8.45% | | | | | |

*Table 5: Optimized parameters for Subject 2*

| Joint | $f_n(Hz)$ | $\zeta$ | $k(N/m)$ | $\alpha_{nl}(m^{-1})$ | Mass $(kg)$ |
|---|---|---|---|---|---|
| Pelvis | 12.504 | 0.0418 | 263.58 | 14.653 | 7.32 |
| Lower torso | 13.874 | 0.1213 | 271.07 | 12.211 | 8.65 |
| Central torso | 5.686 | 0.3851 | 34.48 | 9.769 | 11.98 |
| Upper torso | 10.013 | 0.0418 | 59.53 | 7.327 | 9.32 |
| Neck | 18.140 | 0.1624 | 74.35 | 3.907 | 1.33 |
| Head | 10.633 | 0.1735 | 19.60 | 1.954 | 4.39 |
| Rayleigh $\alpha_r = 1.9041\frac{rad}{s}$, Mass scale =0.992, $nl_{scale} = 0.488$, weighted error= 8.37% | | | | | |

## 8.4 Forward dynamics for subject 1

The idealized parameters for subject 1 were employed to predict the biodynamic response for an unused train-road crossing event. The acceleration profiles, apparent mass, impedance, STHT and injury and discomfort parameters are presented in Figure 3 to Figure 9.

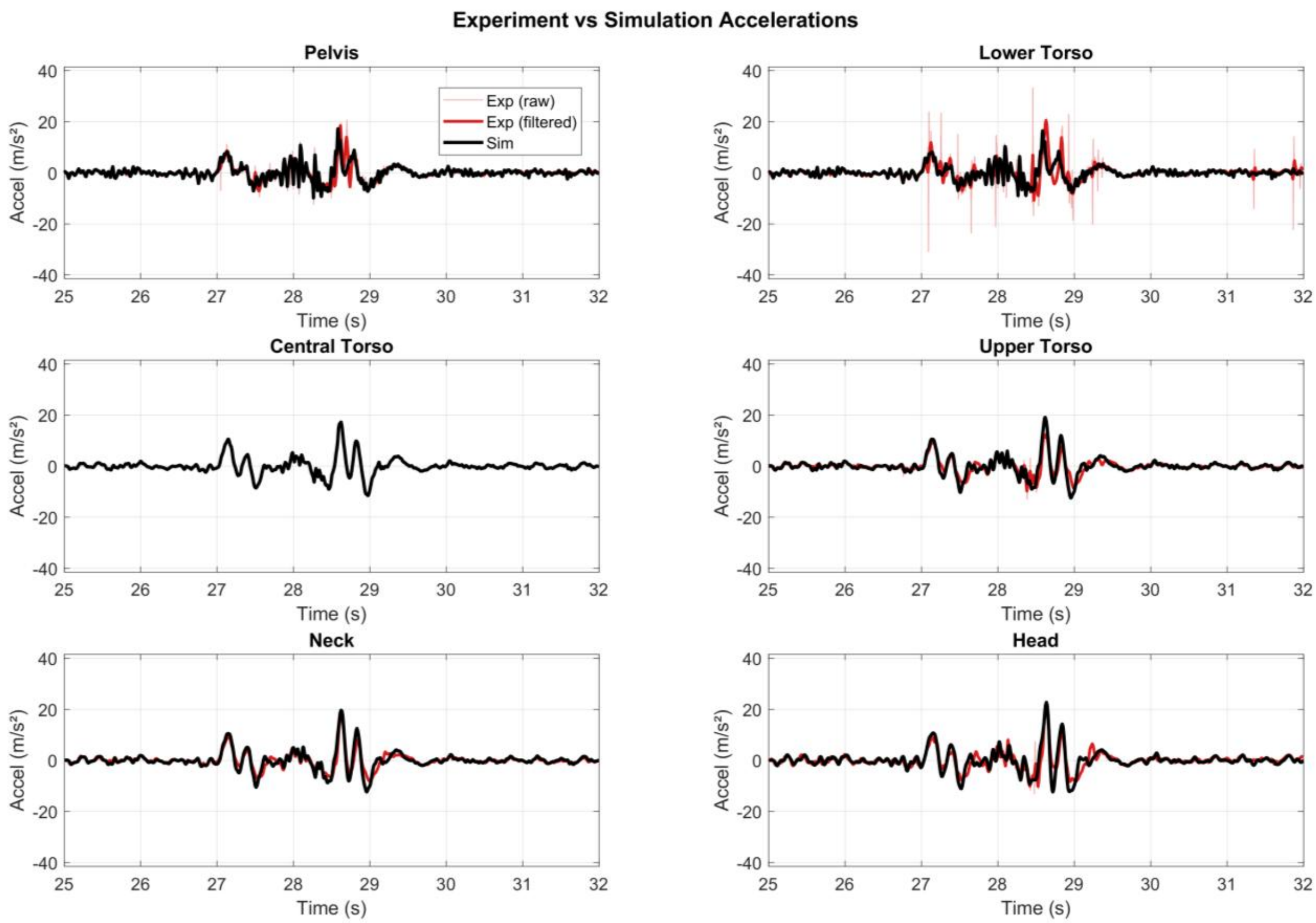

*Figure 3: Subject 1 experimental versus simulated acceleration overlay. The agreement across multiple channels confirms that the model captures the main transient structure well, even though local mismatches remain in the lower torso.*

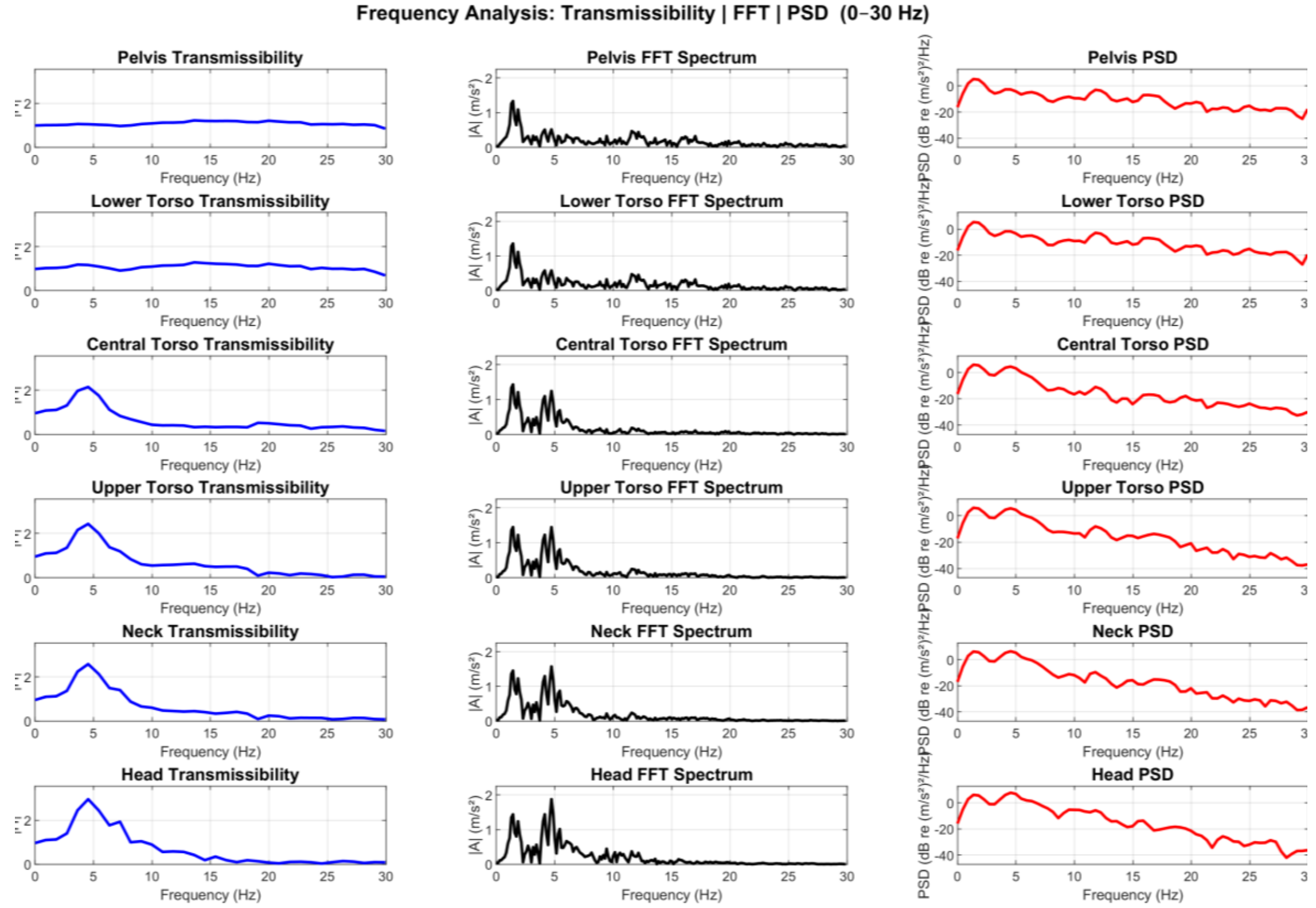

*Figure 4: Subject 1 transmissibility, FFT spectrum, and PSD. These spectral views show the body as a broadband transmission system and help interpret why particular segments accumulate more vibration dose than others and which frequency components dominate the response at each segment.*

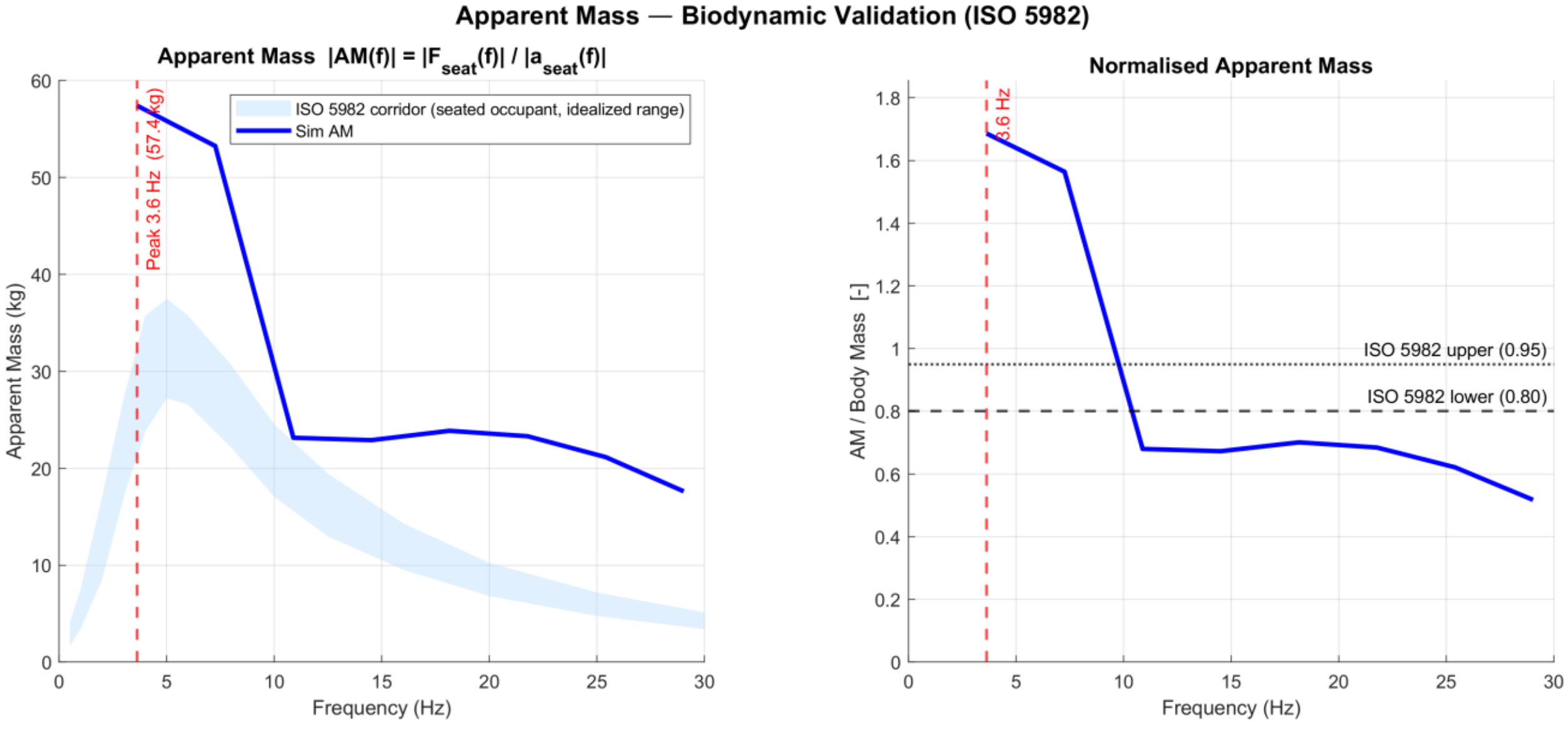

*Figure 5: Subject 1 apparent mass. The reported apparent-mass peak is 57.4 kg at 3.6 Hz, with normalized apparent mass 1.686, above the nominal ISO corridor, indicating stronger dynamic seat-interface coupling than the reference target range.*

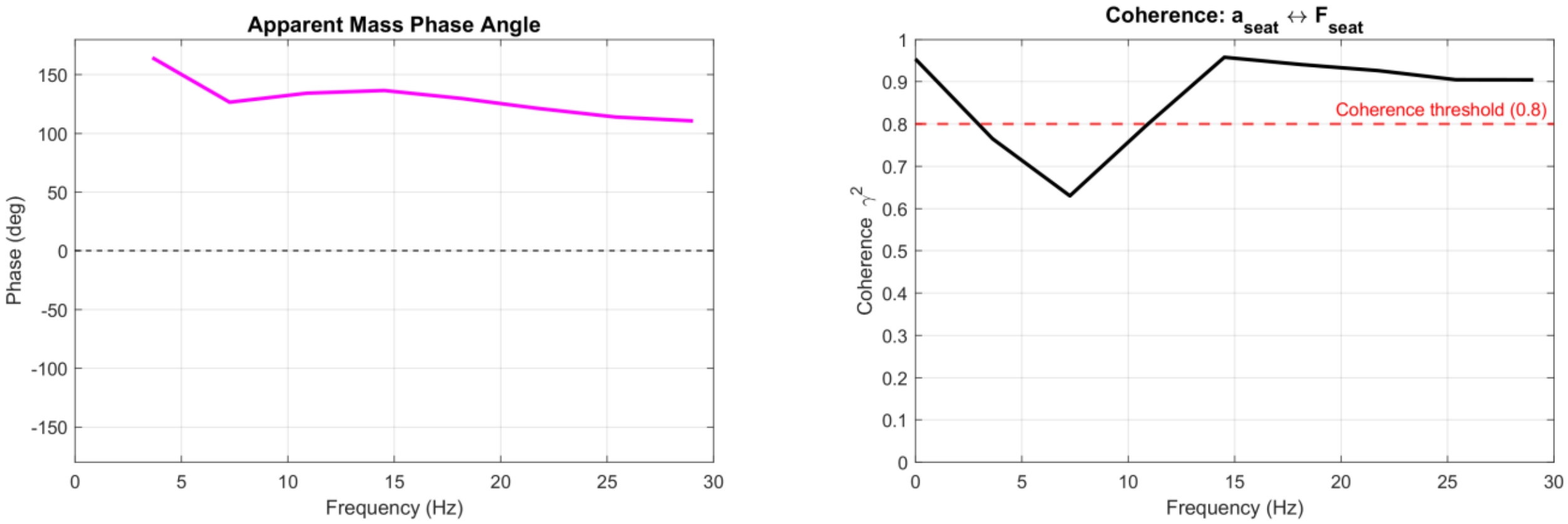

*Figure 6: Subject 1 combined STHT and apparent-mass panel. The figure is useful because it unifies internal transmission behavior and external seat-interface biodynamics in one comparison*

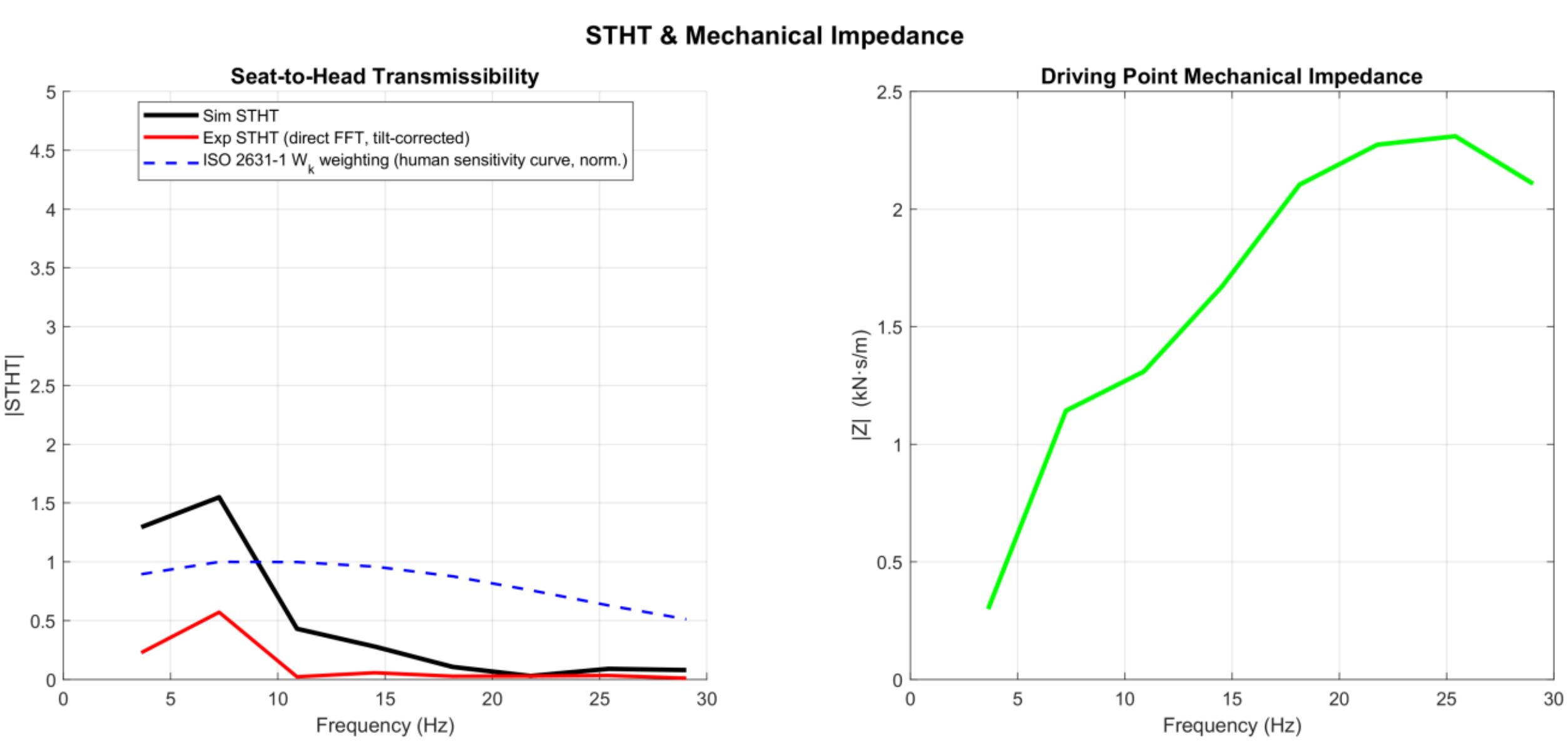

*Figure 7:Subject 1 seat-to-head transmissibility and mechanical impedance. The low-frequency STHT trend helps explain head amplification, while the impedance curve links internal biodynamics to the dynamic load seen at the seat interface.*

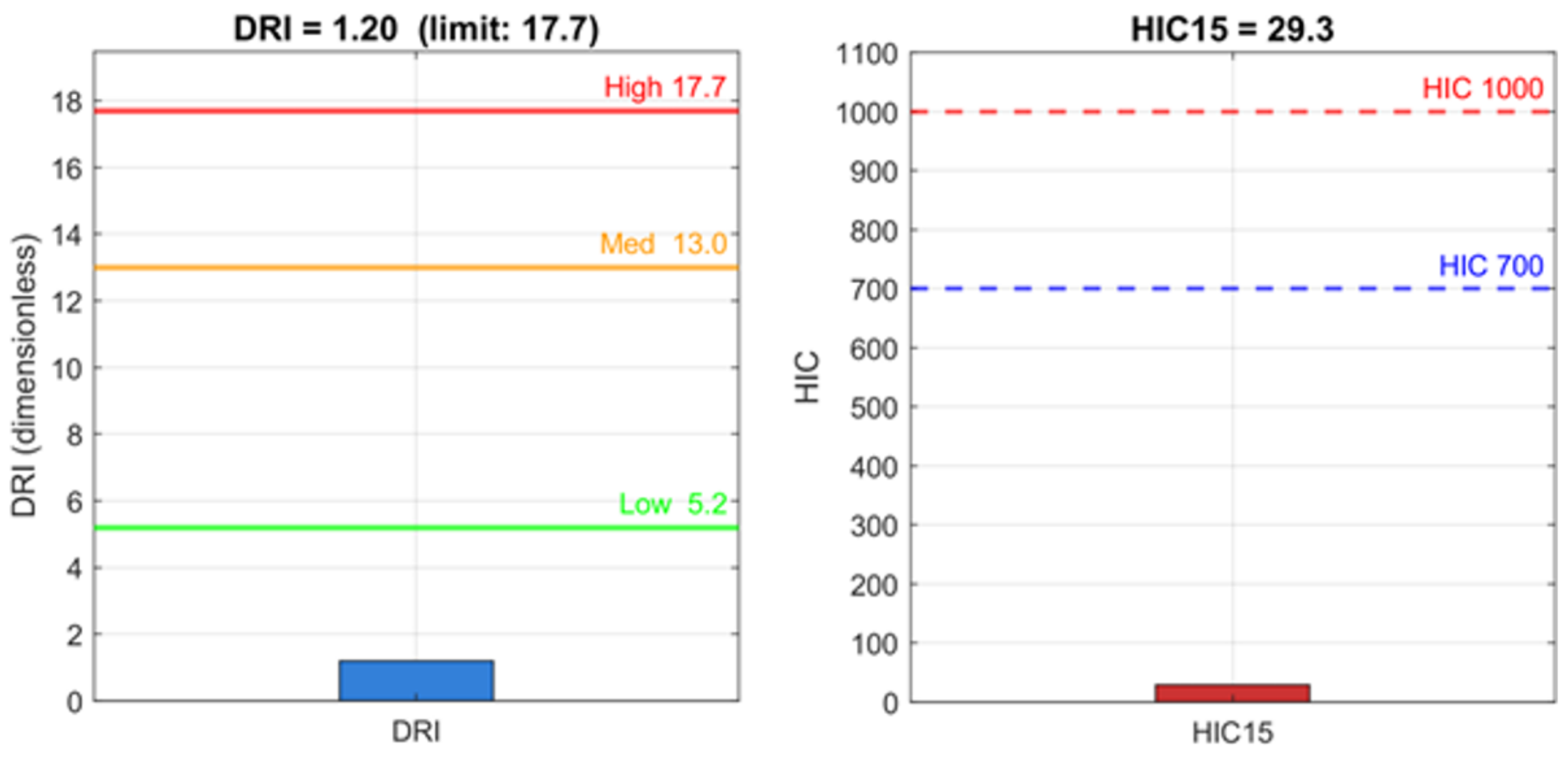

*Figure 8: DRI and* $HIC_{15}$ for subject 1.

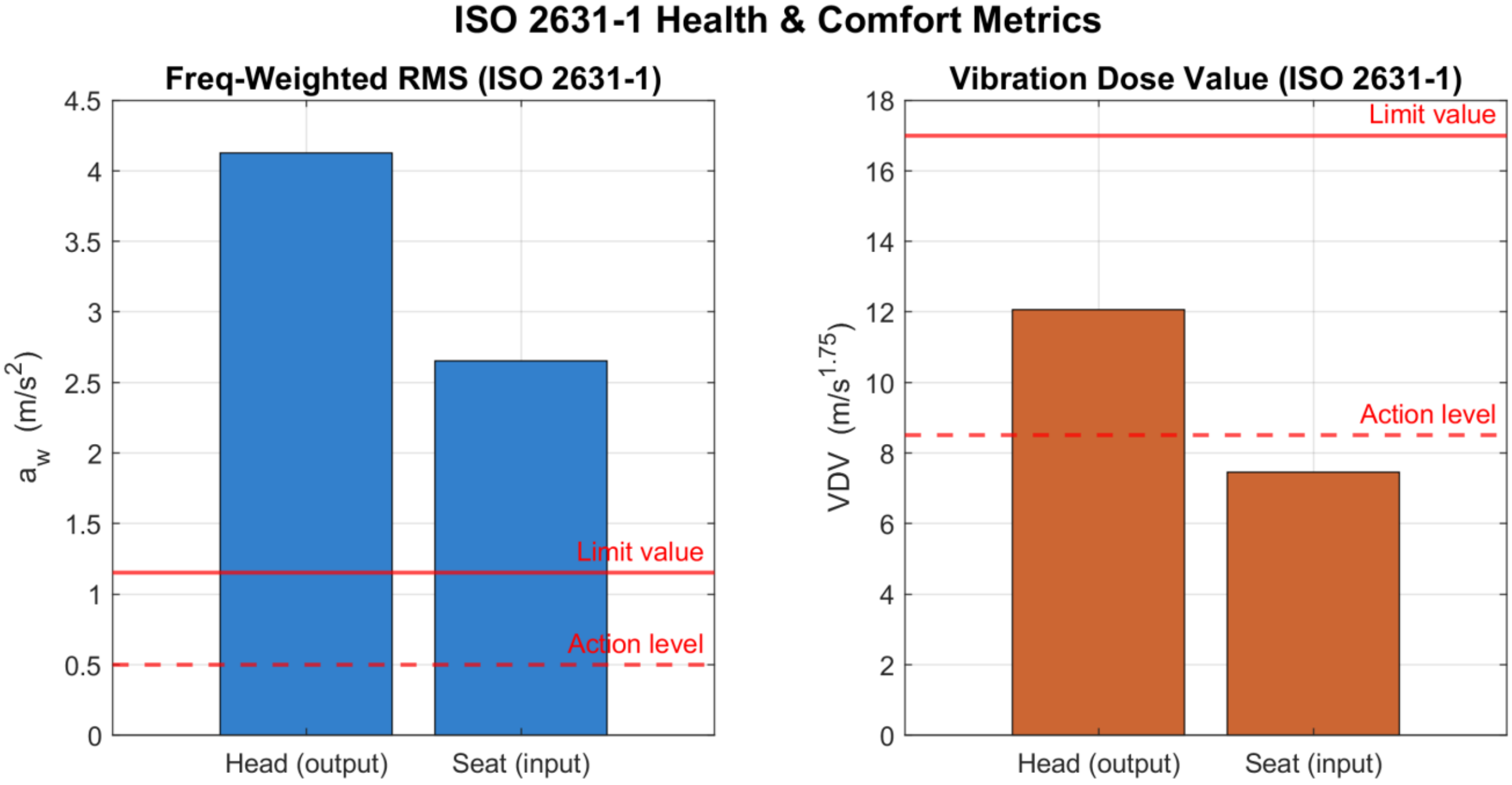

*Figure 9: Subject 1 vibration dose value by segment. This figure shows that exposure burden varies by segment and cannot be inferred from seat input alone.*

## 8.5 Forward dynamics for subject 2

The idealized parameters for subject 2 were employed to predict the biodynamic response for an unused train-road crossing event. The acceleration profiles, apparent mass, impedance, STHT, and injury and discomfort parameters are presented in Figures 10 to Figure 16.

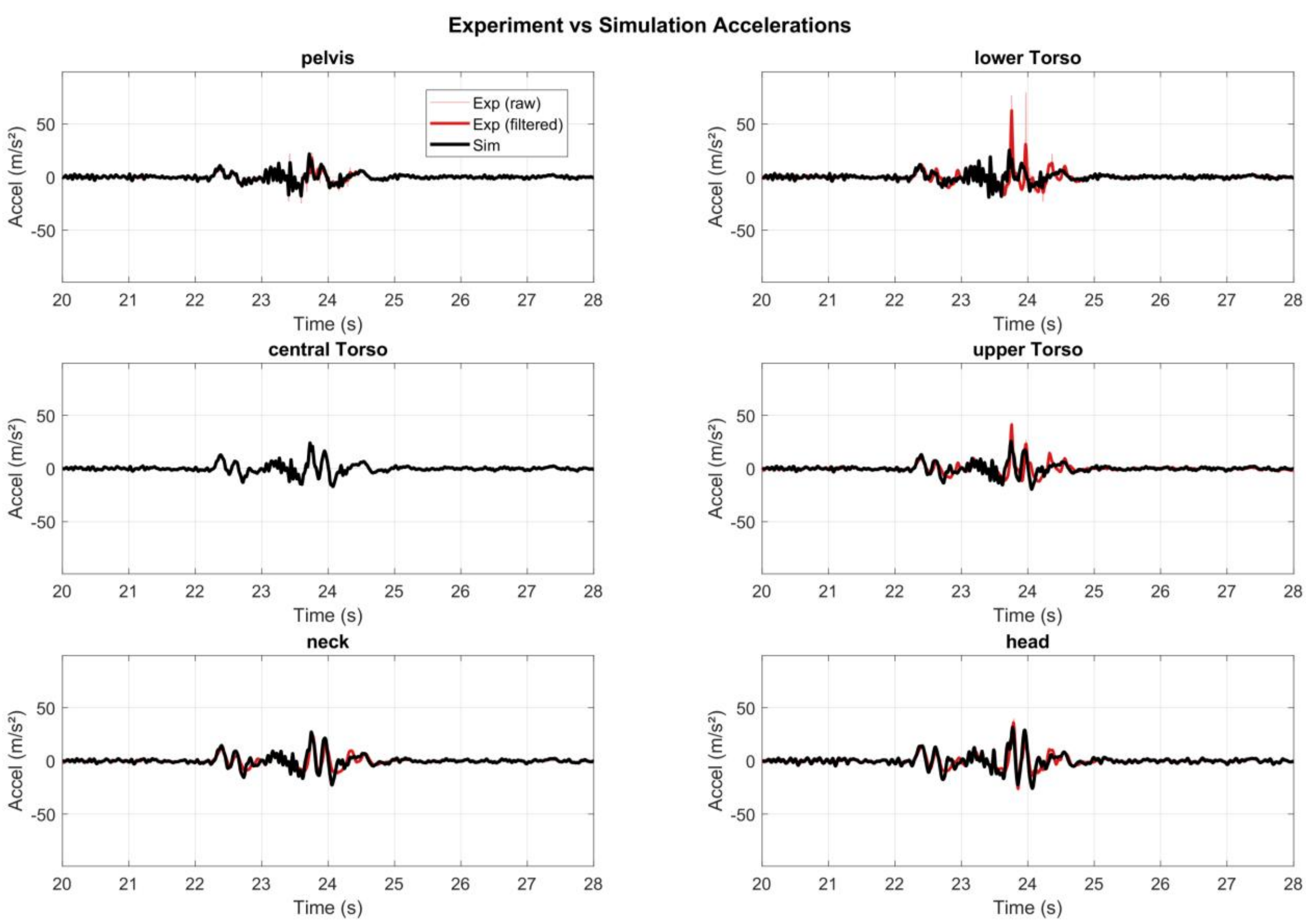

*Figure 10: Subject 2 experimental versus simulated acceleration overlay. The model reproduces the main timing and envelope of the measured responses, although local lower-torso mismatch persists.*

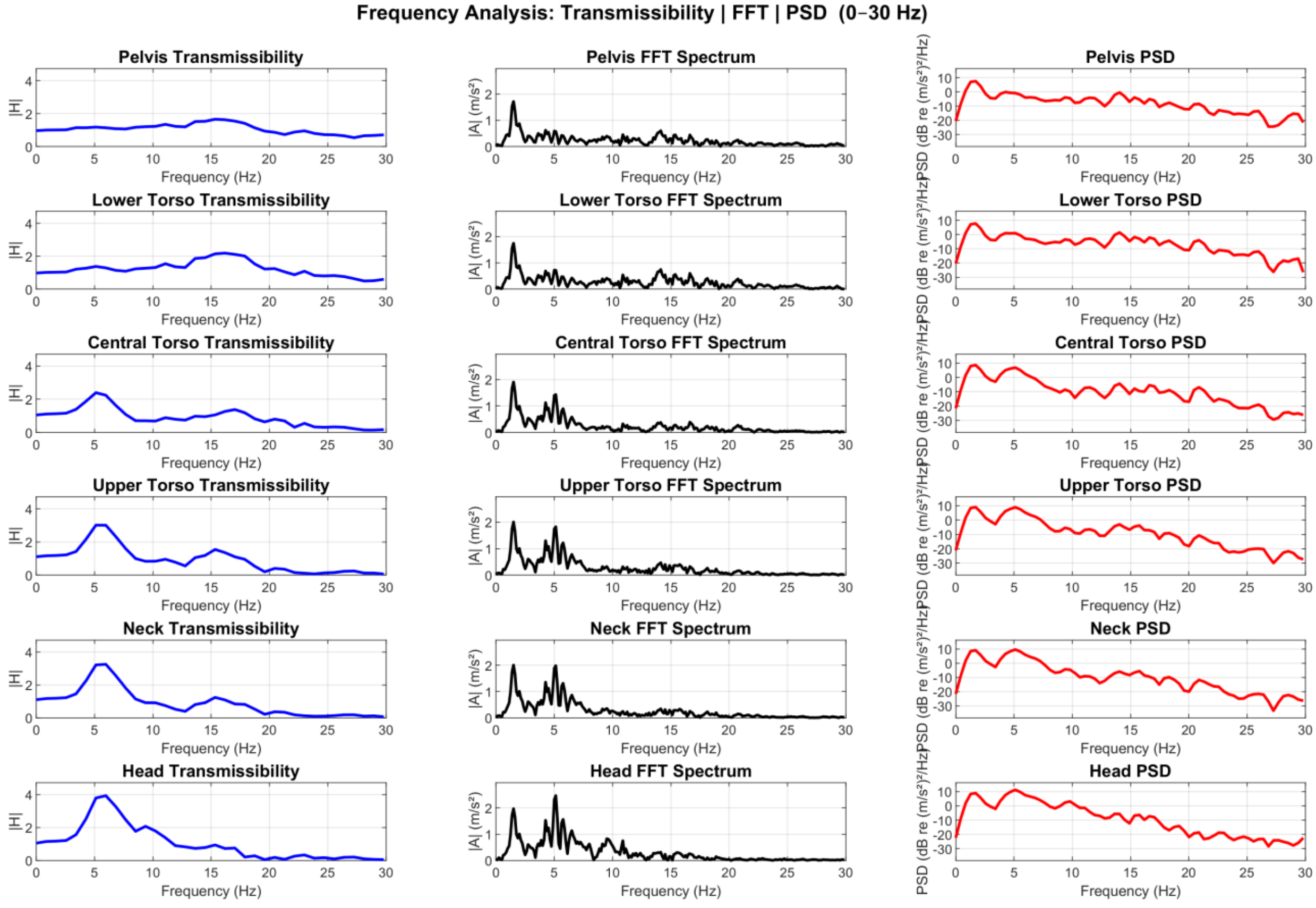

*Figure 11: Subject 2 transmissibility, FFT, and PSD. The figure reveals the broadband structure of the response and helps explain why particular segments accumulate more dose and weighted acceleration.*

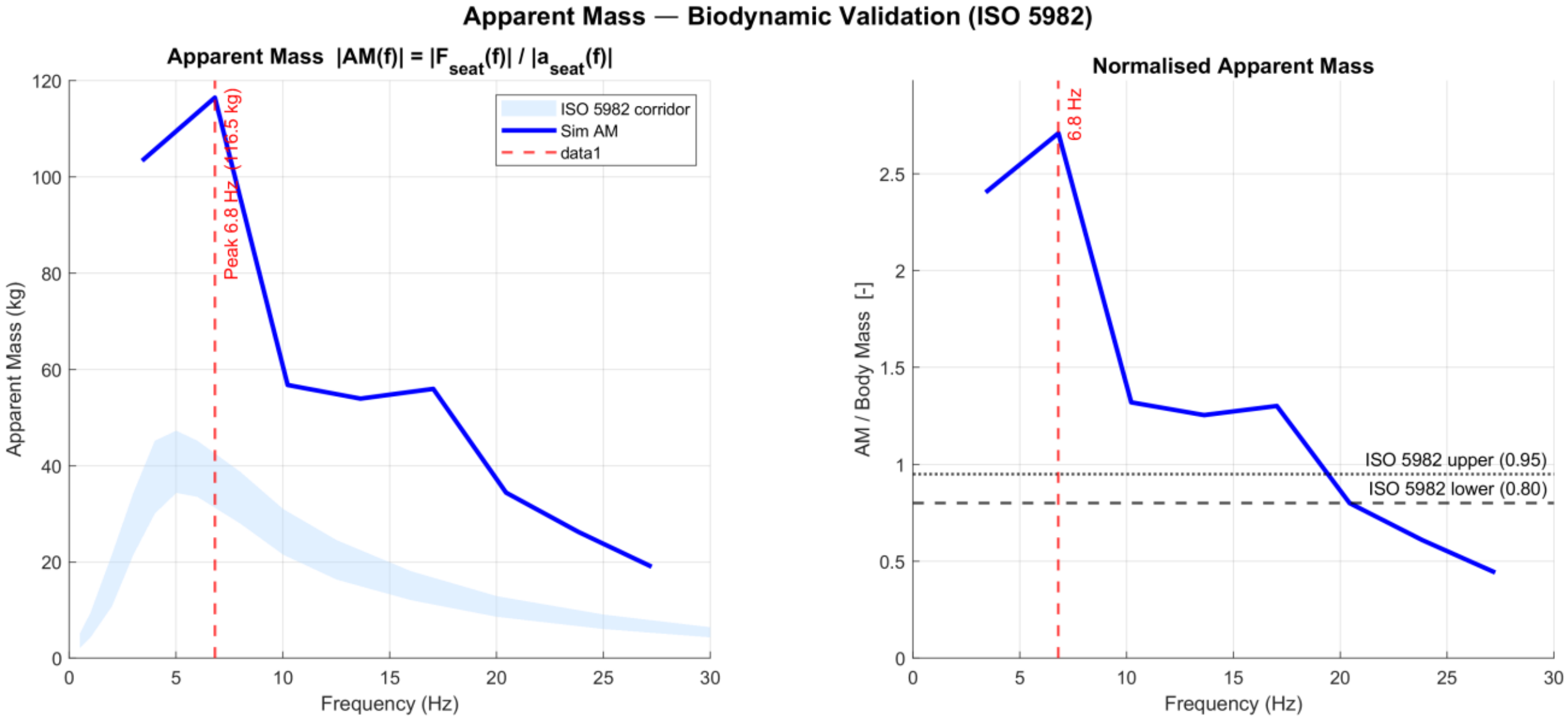

*Figure 12:Subject 2 apparent mass. The reported peak apparent mass is 116.5 kg at 6.8 Hz, with normalized apparent mass 2.709, far above the nominal ISO target range and substantially larger than Subject 1*

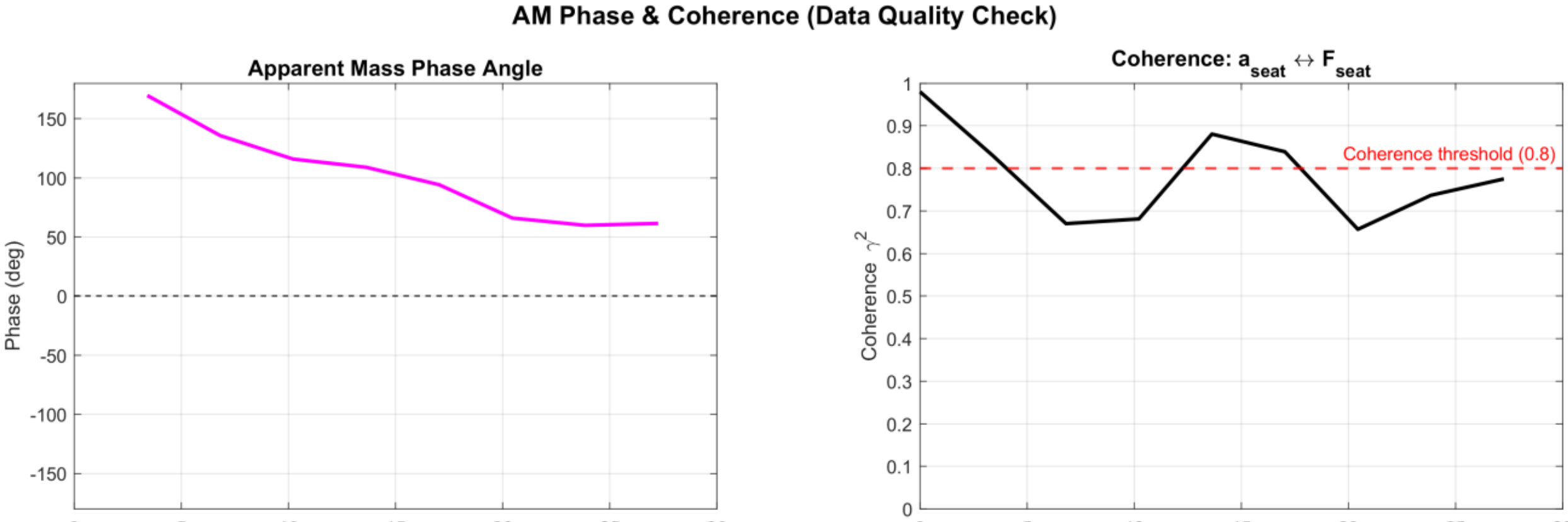

*Figure 13: Subject 2 combined STHT and apparent-mass panel. This figure is especially useful for discussion because it links transmission to the head with driving-point response at the seat.*

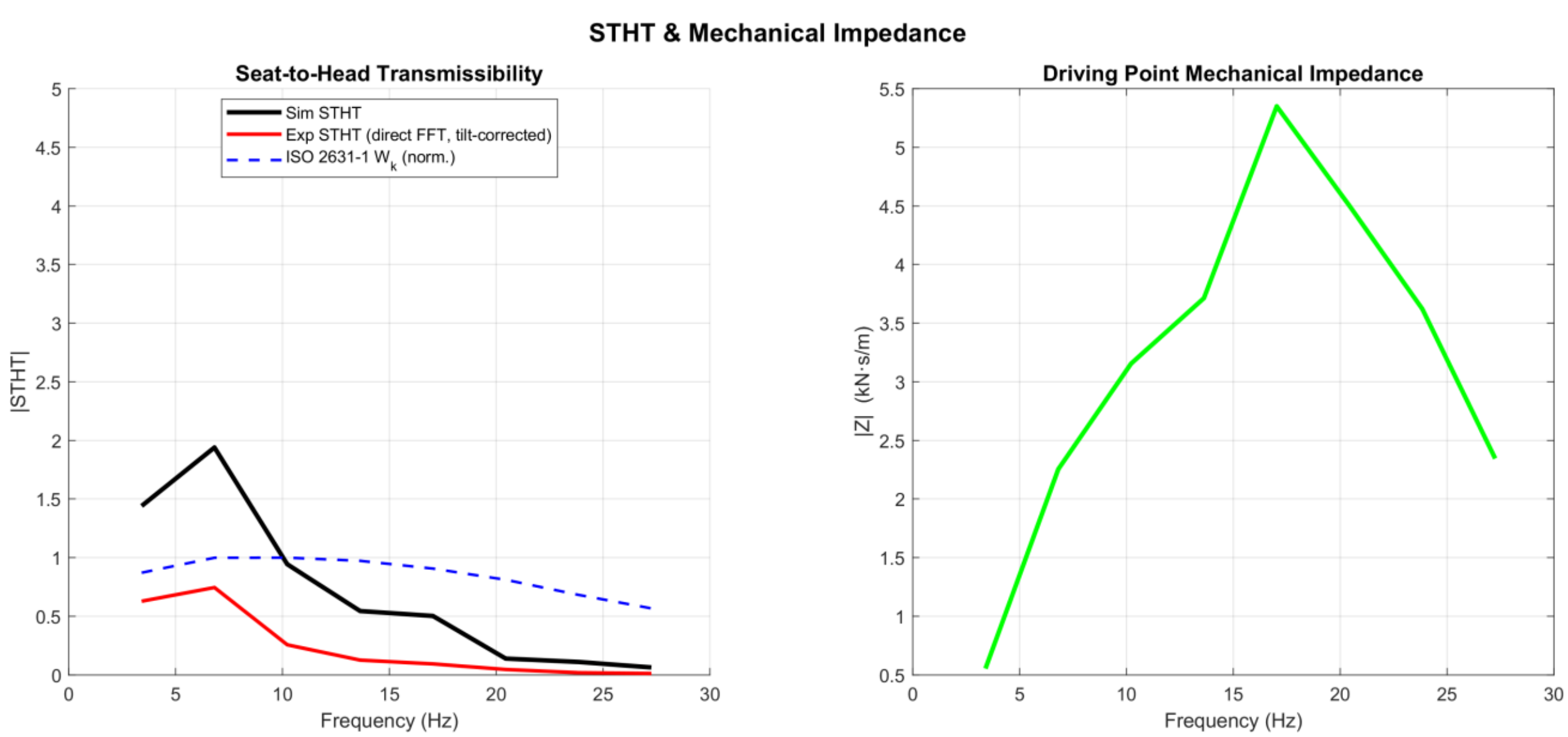

*Figure 14: Subject 2 STHT and mechanical impedance. The figure indicates some low-frequency overprediction of seat-to-head transmissibility, suggesting that the model slightly over-amplifies head motion in the resonant region*

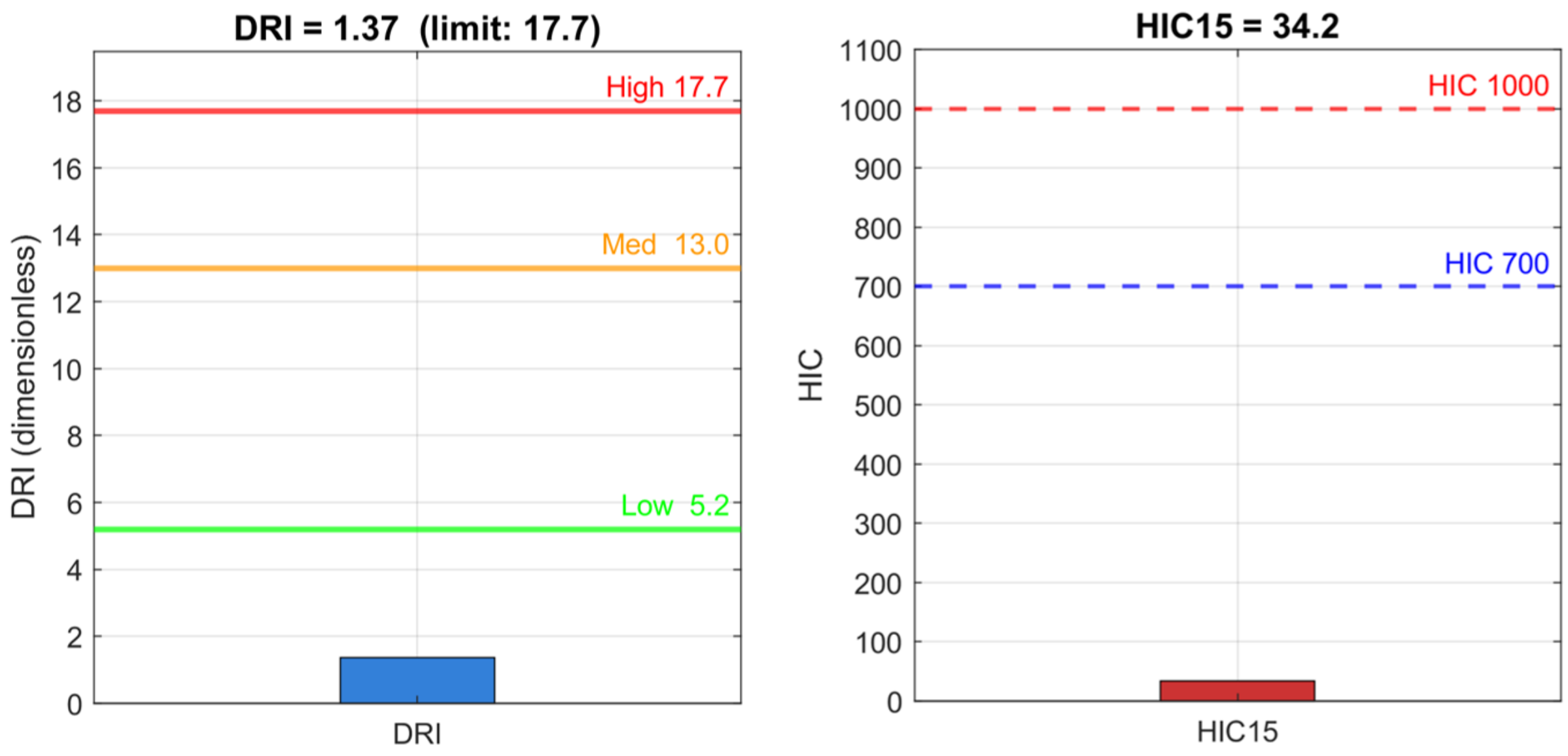

*Figure 15: DRI and $HIC_{15}$ for subject 2.*

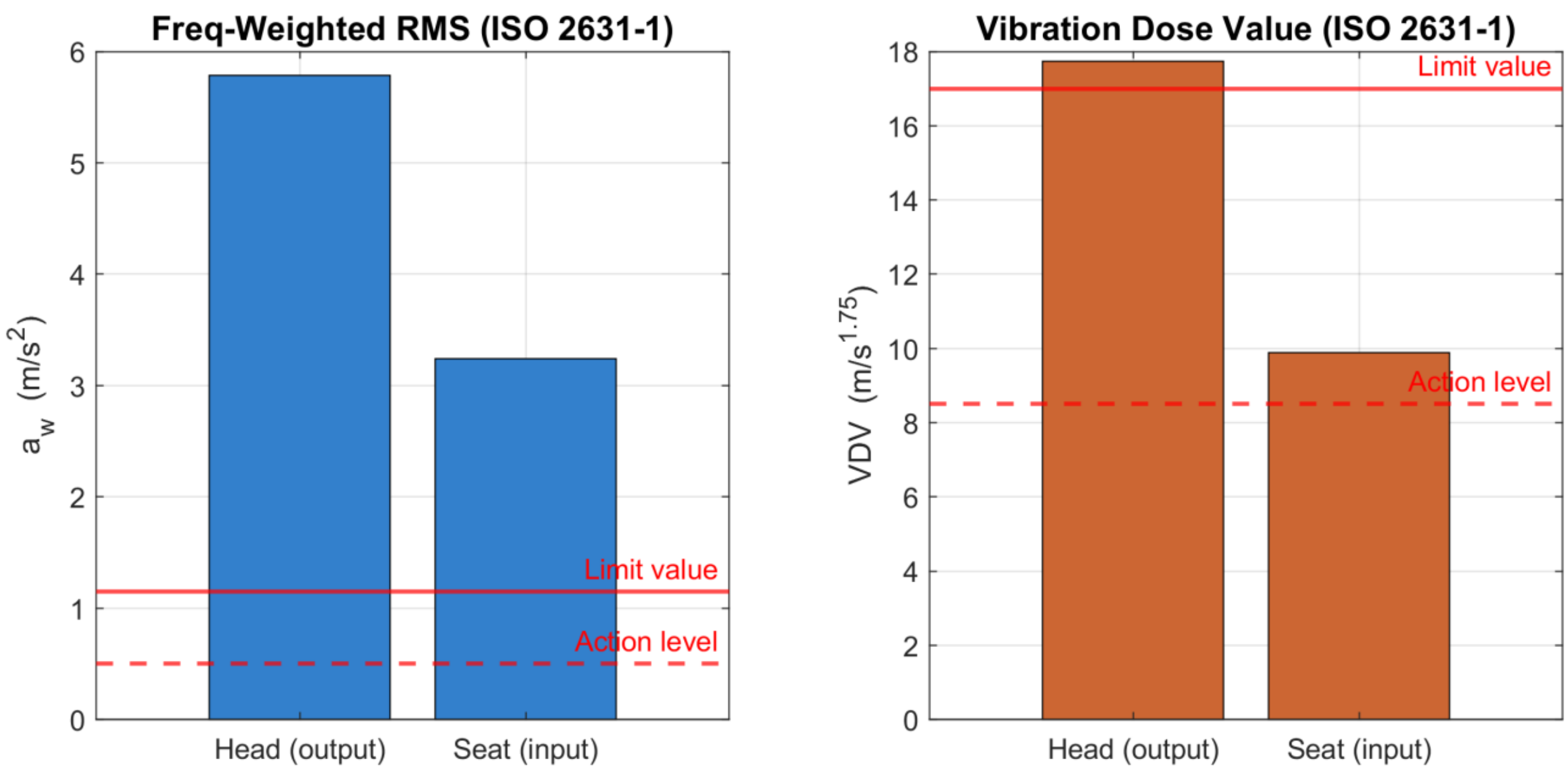

*Figure 16: Subject 2 vibration dose value by segment. The body-wide distribution of VDV confirms that distal and flexible segments carry disproportionate exposure burden.*

# 9 DISCUSSION

In this study, a two-phase optimization framework successfully implemented to identify subject-specific parameters sets for both participants with idealized weighted-mean errors of less than 1-% for both subjects cross three mechanically distinct loading conditions. For subject 1, the weighted mean errors was 8.45% and for subject 2, it was 8.37%. We compared the NRMS between the model's predicted absolute acceleration and the measured IMU signals at five body locations. To contextualize these error, Boileau and Rakheja (Boileau & Rakheja, 1998) reported 20% standard error on magnitude and up to 50% for phase near resonance peak for a four-DOF seated human model. The testing range was 0.1g to 0.2g. In another study, Liang and Chiang (Liang & Chiang, 2006) mentioned 10-15% average error in predicting biodynamic response for a four-DOF model. The testing range was broadband type with maximum acceleration of 0.5g.Toward et al. (Toward & Griffin, 2011) reported about 2-% to 30% variability in the resonance frequency and transmissibility for a testing range of maximum peaks 0.15g. The present six-DOF model therefore achieves better accuracy while simultaneously accommodating three fundamentally different excitation regimes for multiple subjects. On top of that the testing range acceleration was non-stationary random with maximum range up to 10g. This breadth of validation represents a significant advance over prior single-condition and single subject identification studies.

### 9.1 Inter subject comparison and generalizability

The two subjects represent meaningfully different points in the body-mass distribution: Subject 1 at 58.6 kg is at approximately the 25th percentile for adult U.S. males, while Subject 2 at 74.0 kg is closer to the 50th percentile.

Apparent mass reflects how much force the seat must generate to accelerate the body at a given frequency. A larger peak usually indicates stronger resonance, greater energy storage in the torso, and more efficient transmission of vibration into the occupant's body. From the apparent mass figures (Figure 5 and Figure 13), we can see that Subject 2 shows a substantially larger and higher-frequency apparent mass which indicates a stronger resonant coupling, and greater vibration transmission than Subject 1 under the same loading condition. For subject 1, the apparent-mass peak is 57.4 kg at 3.6 Hz, versus 116.5 kg at 6.8 Hz for Subject 2, which means Subject 2's upper body reacts as a much "heavier" dynamic system over its resonant band.

The STHT curves (Figure 7 and Figure 14) indicate the resonant frequencies at which seat vibration is most efficiently transmitted to the head. The differences in peak magnitude and peak location between the two subjects reveal differences in their biodynamic coupling and damping characteristics. The STHT plots show both subjects' STHT peak in roughly the same low-frequency band near 7 Hz. However, the peak STHT for Subject 2 (1.95) is about 26% higher than Subject 1 (1.55) in simulation, which means Subject 2 transmits more seat vibration to the head near resonance. The experimental peak is also higher for Subject 2 (0.75), about 31% higher than Subject 1 (0.57). Subject 1 drops below 1 faster after the resonance, reaching about 0.65 at 10 Hz and about 0.25 near 15 Hz, while Subject 2 is still about 0.95 at 10 Hz and about 0.55 at 15 Hz. That tells you Subject 2 keeps transmitting vibration over a wider frequency range, while Subject 1 attenuates more strongly once past the low-frequency peak. Overall, Subject 2 is more resonant and more strongly coupled, while Subject 1 is more damped and attenuates high-frequency motion more quickly, even though both subjects' STHT peak in roughly the same low-frequency band near 7 Hz for both experimental and simulation data.

The impedance curves (Figure 7 and Figure 14) for both subjects rises with frequency, peaking around 18–20 Hz, which suggests the seated body becomes dynamically harder to move at higher frequencies. The impedance curve for Subject 2 rises to about $5.3\frac{kNs}{m}$, which is more than twice for Subject 1 ($2.3\frac{kNs}{m}$). A higher impedance peak means the system resists motion more strongly at those frequencies, but it does not prevent the resonance peak from being larger. Overall, it shows stronger frequency-dependent dynamics. That is consistent with a resonance-dominated system where low-frequency motion is transmitted more easily, and then the system stiffens and suppresses transmission as frequency increases.

#### 9.1.1 Convergence parameters

The convergence parameters in Table 6 indicates that several parameters are relatively stable across subjects and can therefore be treated as population-based starting values. The natural frequencies $f_n$ of the head, neck, upper torso, central torso, and pelvis all remain below the 30% threshold, which suggests that the linear resonance behavior of these segments is broadly transferable. The mass scale factor $m_{scale}$ also shows very small deviation, reinforcing the idea that global mass adjustment is not strongly subject-dependent in this case.

*Table 6 Convergence Parameters Based on Absolute Percentage Deviation Threshold*

| Parameters | Absolute deviations (%) | Parameters | Absolute deviations (%) |
|---|---|---|---|
| Head $f_n$ | 2.49 | UT $k$ | 3.17 |
| Neck $f_n$ | 13.06 | Pelvis $k$ | 8.89 |
| UT $f_n$ | 19.76 | Head $\zeta$ | 14.98 |
| CT $f_n$ | 12.44 | UT $\zeta$ | 18.68 |
| Pelvis $f_n$ | 14.77 | Mass scale $m_{scale}$ | 0.98 |

The relatively low deviation in $f_n$ does not mean the full mechanical response is identical, but it does indicate that the dominant linear dynamic characteristics are similar between the two subjects. This makes these parameters useful for a population-average model, especially when the goal is to obtain a reasonable initial estimate before subject-specific refinement. Among the convergent variables, the head $f_n$ is the most stable, followed by neck $f_n$, upper torso $f_n$, central torso $f_n$, and pelvis $f_n$, all of which remain within the selected threshold.

### 9.1.2 Divergence parameters

The divergence parameters in Table 7 shows that the nonlinear and damping-related parameters vary much more strongly across subjects, so they should be treated as subject-specific rather than population-average values. The nonlinear stiffness terms $\alpha_{nl}$ for the head, neck, upper torso, and central torso all exceed 100% deviation, which clearly indicates strong inter-subject variability in the nonlinear force-deformation behavior. The damping ratios $\zeta$ for the neck, central torso, lower torso, pelvis, and upper torso also show large deviations, meaning energy dissipation characteristics are not well captured by a single shared value.

*Table 7 Divergence Parameters Based on Absolute Percentage Deviation Threshold*

| Parameters | Absolute deviations (%) | Parameters | Absolute deviations (%) |
|---|---|---|---|

| Neck $\zeta$ | 113.39 | Head $\alpha_{nl}$ | 125.38 |
|---|---|---|---|
| CT $\zeta$ | 40.70 | Neck $\alpha_{nl}$ | 125.35 |
| LT $\zeta$ | 69.55 | UT $\alpha_{nl}$ | 125.38 |
| Pelvis $\zeta$ | 85.14 | CT $\alpha_{nl}$ | 125.39 |
| Head $k$ | 32.64 | $nl_{scale}$ | 124.88 |
| Neck $k$ | 61.44 | $\alpha_r$ | 36.27 |
| CT $k$ | 81.08 | | |
| LT $k$ | 80.15 | | |

The stiffness values $k$ further support this conclusion. While the pelvis and upper torso stiffness values remain below the threshold and are more transferable, the lower torso, central torso, neck, and head stiffness values exhibit large deviations. The Rayleigh coefficient $\alpha_r$ also shows substantial variation, which suggests that even the global proportional damping adjustment is not fully consistent across subjects. Overall, the divergence table shows that nonlinear stiffness, damping, and several segmental stiffness terms require subject-specific identification to accurately represent individual behavior.

Both subjects achieved similar weighted idealization error while retaining clearly different parameter patterns and clearly different comfort severity. Subject 1 exhibited a comparatively softer and more damped overall trunk response, whereas Subject 2 showed stronger stiffness in the central torso and a higher-amplitude response in the metrics tied to head exposure. As a result, Subject 1 had lower weighted injury and comfort burden, lower apparent-mass amplification, and lower motion-sickness accumulation, while Subject 2 showed stronger amplification, larger dose accumulation, and higher sickness indices. This points to different transmission pathways from the seat into the trunk and head for two subjects.

It means that the framework is stable enough to generate subject-specific consensus models, but still sensitive enough to preserve subject-specific biomechanics. From a biomechanics perspective, that kind of inter-subject spread is exactly why population-average models are limited. A model that fits the average of the two subjects would likely miss the upper-bound response of Subject 2 and overstate the comfort burden of Subject 1.

### 9.2 Parameter variability across loading cases

The spread across case-wise parameters in Table 8 indicates that the identified biodynamic properties are partly load-dependent, and the model responds to differences in input waveform shape, amplitude content, and transient timing. It also shows the importance of amplitude-dependent stiffening correction in subject-specific modeling.

*Table 8: Coefficient of Variation (CV = σ/μ).*

| CV Parameter | Pelvis | LT | CT | UT | Neck | Head |
|---|---|---|---|---|---|---|
| Sub01: CV($f^h$) % | 6.9 | 56.3 | 84.8 | 19.6 | 28.5 | 10.1 |
| Sub02: CV($f^h$) % | 19.1 | 53.7 | 73.1 | 35.4 | 9.6 | 11.9 |
| Sub01: CV(ζ) % | 69.3 | 1.1 | 71.6 | 14.7 | 57.6 | 71.0 |
| Sub02: CV(ζ) % | 19.0 | 61.0 | 9.3 | 19.2 | 55.9 | 68.9 |
| Sub01: $\alpha_r$ | 1.3 | | | | | |
| Sub02: $\alpha_r$ | 37.4 | | | | | |
| Sub01: $nl_{scale}$ | 32.8 | | | | | |
| Sub02: $nl_{scale}$ | 84.2 | | | | | |
| Sub01: $m_{scale}$ | 0.6 | | | | | |
| Sub02: $m_{scale}$ | 9.8 | | | | | |

Coefficient in variations in Table 8 shows how much parameters deviate because of load variations. For both subjects, the consensus idealized parameters captured the common structure of the three cases while retaining biologically meaningful variability. The lower-torso and central-torso frequency and stiffness parameters were more variable than the head and neck parameters, whereas upper body damping parameters are more variable than lower body. That pattern is useful because it implies the optimizer is localizing uncertainty in the segments that are actually hardest to identify from the available data. For future work, the case-to-case spread could be treated as a confidence envelope around the reported subject-specific parameters, which would make the model more useful for prediction under unfamiliar loading conditions.

### 9.3 Injury risk observations

The strongest physiological conclusion is the separation between injury and discomfort. $HIC_{15}$, $N_{ij}$ and DRI remained below risk threshold in both subjects. The comfort and sickness metrices tell the broader story. The weighted RMS, and VDV were all elevated, especially for subject 2. This means the train events studied here are better interpreted as high-discomfort, high-exposure events rather than acute injury events, which makes the model especially relevant for land, air, and water vehicles ride quality and human vibration exposure assessment.

### 9.4 Limitation

Although this work shows significantly improved accuracy in identifying subject-specific biodynamics parameters, there are several limitations with the present work. First, the current model is constrained to the vertical axis. Cars and marine operations involve significant fore-aft and lateral seat accelerations, particularly during cornering or wave-oblique encounters. Extending the model to three translational and three rotational degrees of freedom for each segment would enable the model to predict biodynamic responses and injury probability in a broad range of real-world scenarios. Second, the inter- and intra-individual studies are preliminary as only two subjects were recruited. Recruiting at least eight to ten subjects spanning a wider range of body mass, stature, and seated posture would provide us better accuracy and more insights about biodynamic responses. Third, the pelvis segment captures both pelvis and seat properties. As a result, the model is vehicle-specific. Also, the repeated underprediction of the lower torso by approximately 20% in both subjects is an important structural limitation identified in the present study. Persistent underprediction there suggests that the intermediate trunk may require richer compliance, improved parameter observability, or an additional segment if the goal is higher-fidelity prediction across all conditions. These findings identify a concrete path for future model refinement. Although the model and optimizer have some limitations, they provide important subject-specific accelerations and biodynamic response with high accuracy. Also, it shows convergent and divergent parameters that can help us understand which parameters should be subject-specific and which parameters can be population-based. These insights are useful in protective equipment and vehicle suspension design.

# 10 CONCLUSIONS

A subject-specific six-DOF nonlinear biodynamics model of the human upper body has been developed and validated for two subjects under three distinct loading conditions. A novel two-phase optimization strategy and parameter idealization with a strain stiffening mechanism are used to achieve segment-level acceleration error below 10%. In the two-phase optimization, PSO provides global search, followed by bounded Nelder-Mead time-domain ODE refinement of 15 parameters. Finally, an error-weighted geometric mean in log space yields robust consensus parameters that are independent of the calibration loading, enabling cross-scenario forward predictions without re-identifying the parameters. Idealized parameters are successfully applied to predict upper body vibration response under experimental impulse loading conditions. Comprehensive injury assessments for both subjects show that train crossing-type short, impulsive loadings pose a low risk of acute injury but a high risk of discomfort and motion sickness. Future work should focus on improving the lower-torso observability and expanding the subject pool. Overall, the model provides a strong framework for predicting subject-specific seated human biodynamic responses, and the results demonstrate the practical utility of subject-specific idealized biodynamic models for occupational health assessment and protective system design in maritime and ground vehicle applications.

# 11 AUTHORS CONTRIBUTION STATEMENT:

R.Z. and N.I.H.R. developed the theoretical and computational models and performed the optimization. A.Z.R. and M.I.H. conducted experiments, data collection, and data analysis. A.A. designed the study, acquired the funding, and revised the manuscript. All authors reviewed and approved the final version of the manuscript.

# 12 ACKNOWLEDGEMENT:

A.A. also acknowledges the financial support from Office of Naval Research (ONR)'s DURIP program (N00014-23-1-2232) and Force Health Protection (FHP) program (through the Award # ONR: N00014-21-1-2051 and ONR: N00014-21-1-2043: Dr. Timothy Bentley, Program Manager).